\documentclass[
reprint,
amsmath,amssymb,
aps,
pra,
]{revtex4-2}

\usepackage{graphicx, color}
\usepackage{dcolumn}
\usepackage{bm}
\usepackage{here}

\begin{document}

\title{Limitation of single-repumping schemes for laser cooling of Sr atoms}

\author{Naohiro Okamoto}
\email{okamoton@g.ecc.u-tokyo.ac.jp}
\author{Takatoshi Aoki}
\author{Yoshio Torii}
\email{ytorii@phys.c.u-tokyo.ac.jp}

\affiliation{Institute of Physics, The University of Tokyo, 3-8-1 Komaba, Megro-ku, Tokyo 153-8902, Japan}
\date{\today}

\begin{abstract}
We explore the efficacy of two single-repumping schemes, $5s5p \,{}^3P_2 - 5p^2 \,{}^3P_2$ ($481\,\mathrm{nm}$) and $5s5p \,{}^3P_2 - 5s5d \,{}^3D_2$ ($497\,\mathrm{nm}$), for a magneto-optical trap (MOT) of Sr atoms.
We reveal that the enhancement in the MOT lifetime is limited to $26.9(2)$ for any single-repumping scheme.
Our investigation indicates that the primary decay path from the $5s5p \,{}^1P_1$ state to the $5s5p \,{}^3P_0$ state proceeds via the $5s4d \,{}^3D_1$ state, rather than through the upper states accessed by the single-repumping lasers.
We estimate that the branching ratio for the $5s5p \,{}^1P_1 \to 5s4d \,{}^3D_1 \to 5s5p \,{}^3P_0$ decay path is $1:3.9 \times 10^6$ and the decay rate for the transition from the $5s5p \,{}^1P_1$ state to the $5s4d \,{}^3D_1$ state is $83(32)\,\mathrm{s^{-1}}$.
This outcome underscores the limitation on atom number in the MOT for long loading times ($\gtrsim 1 \,\mathrm{s}$) when employing a single-repumping scheme.
These findings will contribute to the construction of field-deployable optical lattice clocks.

\end{abstract}

\maketitle

\section{introduction}
The electronic structure of alkaline-earth-metal (-like) atoms is characterized by the presence of long-lived metastable states and ultra-narrow transitions, offering a diverse array of applications such as precision measurements{\,}\cite{S.L.Campbell2017, E.Oelker2019, T.L.Nicholson2015, W.F.McGrew2018, S.M.Brewer2019, T.Bothwell2019, N.Nemitz2016, BACONcolab2021}, tests of special relativity{\,}\cite{P.Delva2017}, detection of gravitational redshift{\,}\cite{M.Takamoto2020, T.Bothwell2022, X.Zheng2023}, quantum simulation{\,}\cite{S.Kolkowits2017}, quantum information{\,}\cite{M.A.Norcia2019} and search for fundamental physics{\,}\cite{M.S.Safronova2018}, including gravitational wave detection {\,}\cite{S.Kolkowits2016,M.Abe2021} and search for dark matter{\,}\cite{M.Abe2021, T.Kobayashi2022}.
Particularly, optical lattice clocks utilizing Sr have been intensively studied as promising candidates for redefining the second{\,}\cite{N.Dimarcq2024}.

In a standard magneto-optical trap (MOT) of Sr atoms, the $5s^2 \,{}^1S_0 - 5s5p \,{}^1P_1$ transition ($461\,\mathrm{nm}$) is typically employed. 
However, this transition is not completely closed; the atoms can decay from the $5s5p \,{}^1P_1$ state to the $5s4d \,{}^1D_2$ state with a branching ratio of $1:50{\,}000${\,}\cite{L.R.Hunter1986, X.Xu2003}. 
Subsequently, atoms in the $5s4d \,{}^1D_2$ state further decay to the $5s5p \,{}^3P_2$ and $5s5p \,{}^3P_1$ states at a ratio of 2:1{\,}\cite{C.W.Bauschlicher1985, X.Xu2003}. 
Those atoms decaying to the $5s5p \,{}^3P_1$ state return to the $5s^2 \,{}^1S_0$ state at a rate of $4.7\times10^4\,\mathrm{s^{-1}}${\,}\cite{R.Drozdowski1997} and are subsequently recaptured in the MOT. 
Conversely, atoms decaying to the $5s5p \,{}^3P_2$ state leak out of the trap because the lifetime of the $5s5p \,{}^3P_2$ state is approximately $10^3\,\mathrm{s}${\,}\cite{A.Derevianko2001, M.Yasuda2004}.
Therefore, an additional laser repumping the atoms in the ${}^3P_2$ state to the $5s5p \,{}^3P_1$ state is required. 

Earlier experiments utilized the $5s5p \,{}^3P_2 - 5s6s \,{}^3S_1$ ($707\,\mathrm{nm}$) transition for repumping the atoms in the $5s5p \,{}^3P_2$ state. 
However, atoms excited to the $5s6s \,{}^3S_1$ state can decay to the long-lived $5s5p \,{}^3P_0$ state{\,}\cite{A.V.Taichenachev2006, Z.W.Barber2006}, necessitating another laser at the $5s5p \,{}^3P_0 - 5s6s \,{}^3S_1$ ($679\,\mathrm{nm}$) transition{\,}\cite{K.R.Vogel1999}. 

Recently, single-repumping schemes have also been demonstrated, specifically using $5s5p \,{}^3P_2 - 5s5d \,{}^3D_2$ ($497\,\mathrm{nm}$){\,}\cite{N.Poli2005}, $5s5p \,{}^3P_2 - 5s6d \,{}^3D_2$ ($403\,\mathrm{nm}$){\,}\cite{S.Stellmer2014}, $5s5p \,{}^3P_2 - 5p^2 \,{}^3P_2$ ($481\,\mathrm{nm}$){\,}\cite{F.Hu2019}, and $5s5p \,{}^3P_2 - 5s4d \,{}^3D_2$ ($3012 \,\mathrm{nm}$){\,}\cite{P.G.Mickelson2009}.
If the upper state of the repumping transition had a negligible decay rate to the $5s5p \,{}^3P_0$ state, the single-repumping scheme would suffice. 
However, all previous studies utilizing a single-repumping scheme have reported that the trap lifetime ($< 1\,\mathrm{s}$) was much shorter than the vacuum-limited value ($\sim 10\,\mathrm{s}$). 

For a MOT of Yb atoms using the $6s^2 \,{}^1S_0 - 6s6p \,{}^1P_1$ ($399\,\mathrm{nm}$) transition, the dominant loss channel is the decay path of $6s6p \,{}^1P_1 \to 6s5d \,{}^3D_1 \to 6s6p \,{}^3P_0$, limiting the trap lifetime to $\sim 1\,\mathrm{s}${\,}\cite{S.G.Porsev1999, J.W.Cho2012}. 
However, for a Sr MOT, the decay path of $5s5p \,{}^1P_1 \to 5s4d \,{}^3D_1 \to 5s5p \,{}^3P_0$ has rarely been discussed {\,}\cite{S.Stellmer2014}.
To our knowledge, there has been no measurement of the branching ratio for the transition from the $5s5p \,{}^1P_1$ state (via the $5s4d \,{}^3D_1$ state) to the $5s5p \,{}^3P_0$ state.

In this study, we investigate the performance of a MOT of $\mathrm{{}^{88} Sr}$ atoms for two single-repumping schemes: $5s5p \,{}^3P_2 - 5p^2 \,{}^3P_2$ ($481\,\mathrm{nm}$) and $5s5p \,{}^3P_2 - 5s5d \,{}^3D_2$ ($497\,\mathrm{nm}$).
Our findings reveal that the enhancement in the MOT lifetime is $26.9(2)$, irrespective of both the single-repumping transition and the trapped atom density.
This outcome indicates that the dominant decay path from the $5s5p \,{}^1P_1$ state to the $5s5p \,{}^3P_0$ state is via the $5s4d \,{}^3D_1$ state ($5s5p \,{}^1P_1 \to 5s4d \,{}^3D_1 \to 5s5p \,{}^3P_0$). 
Additionally, we estimate that the branching ratio for the transition from the $5s5p \,{}^1P_1$ state (via the $5s4d \,{}^3D_1$ state) to the $5s5p \,{}^3P_0$ state is $1:3.9 \times 10^6$ and the decay rate for the transition from the $5s5p \,{}^1P_1$ state to the $5s4d \,{}^3D_1$ state is $83(32)\,\mathrm{s^{-1}}$.
These findings underscore that, when a long loading time ($\gtrsim 1\,\mathrm{s}$) is necessary, the atom number in the MOT is significantly limited for single-repumping schemes. 
This revelation will aid in the development of field-deployable optical lattice clocks.

\section{experimental setup}
\begin{figure}
	\begin{center}
		\includegraphics[width=86mm]{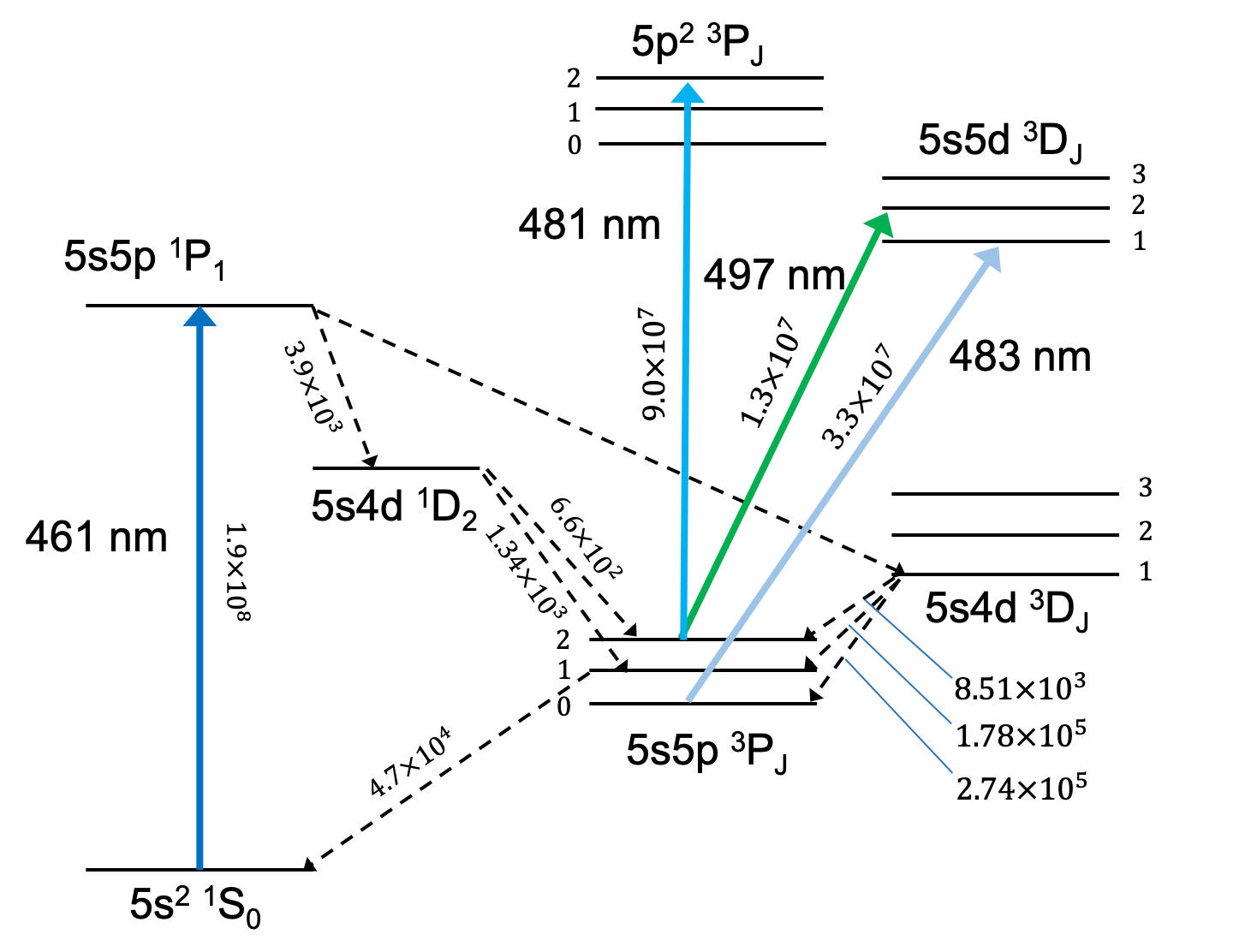}
		\caption{Relevant energy levels for Sr with transition decay rates (in $\mathrm{s^{-1}}$). Solid arrows indicate transitions utilized in our study whereas dashed lines represent relevant decay paths. See Appendix \ref {tables} for references on decay rates.}
		\label{Sr_energy_levels}
	\end{center}
\end{figure}

Figure \ref{Sr_energy_levels} shows the energy levels and decay rates of Sr relevant to our investigation.
The laser for trapping drives the $5s^2 \,{}^1S_0 - 5s5p \,{}^1P_1$ ($461\,\mathrm{nm}$) transition.
Additional lasers are employed to address specific transitions for repumping:
the $5s5p \,{}^3P_2 - 5p^2 \,{}^3P_2$ transition at $481\,\mathrm{nm}$, and
the $5s5p \,{}^3P_2 - 5s5d \,{}^3D_2$ transition at $497\,\mathrm{nm}$ to compare the two ${}^3P_2$ repumping schemes, and
the $5s5p \,{}^3P_0 - 5s5d \,{}^3D_1$ transition at $483\,\mathrm{nm}$ for ${}^3P_0$ repumping.
All the lasers are homemade external-cavity diode lasers (ECDL).

The MOT is composed of three retroreflected beams at $461\,\mathrm{nm}$ with a diameter of 18 mm and a combined power of $65\,\mathrm{mW}$ yielding a total intensity at the MOT position of $48\,\mathrm{mW/cm^2}$, estimated from MOT lifetime measurements (see Appendix \ref{estimation_f}).
The axial magnetic field gradient of the MOT coils is $50\,\mathrm{G/cm}$. 
The repumping beams, each with a power of a few $\mathrm{mW}$ and a diameter of about $5\,\mathrm{mm}$, have sufficient intensities; halving their power does not degrade the MOT performance.
The frequencies of these lasers are stabilized using birefringent atomic vapor laser lock (BAVLL) employing a hollow cathode lamp{\,}\cite{T.Sato2022}.
The detuning of the trapping light is adjusted by introducing an offset to the BAVLL signal, while the frequencies of all the repumping lasers are locked at resonance.

The details of the vacuum system will be described elsewhere.
In summary, a compact oven with capillaries, following a design inspired by Ref.{\,}\cite{M.Schioppo2012}, is utilized for loading atoms in the MOT.
The atoms are loaded in the MOT directly from a thermal atomic beam derived from the oven as demonstrated for Li{\,}\cite{B.P.Anderson1994} and Ca{\,}\cite{C.W.Oates1999}.
The atoms are trapped in a glass cell ($25\,\mathrm{mm}\,\times\,25\,\mathrm{mm}\,\times\,100\,\mathrm{mm}$), and the entire vacuum system is evacuated by a single 55-l/s ion pump.
The distance between the oven and the MOT region is $37\,\mathrm{cm}$.
The oven temperature is set to $335\,\mathrm{{}^\circ C}$, resulting in a MOT loading rate of $5.0 \times 10^5\,\mathrm{atoms/s}$ and a vacuum pressure of $\sim 1 \times 10^{-10}\,\mathrm{Torr}$. 
Under these conditions, with both the $5s5p \,{}^3P_2$ and $5s5p \,{}^3P_0$ repumping laser beams on, the number of trapped atoms is $\sim 2 \times 10^{6}$ as estimated by measuring fluorescence using a photodiode.
The observed decay of trapped atoms, after shutting off the atomic beam, exhibits a lifetime of $15\,\mathrm{s}$, which is presumably limited by background gas collisions.

\section{Results and discussion}
\begin{figure}
	\begin{center}
		\includegraphics[width=86mm]{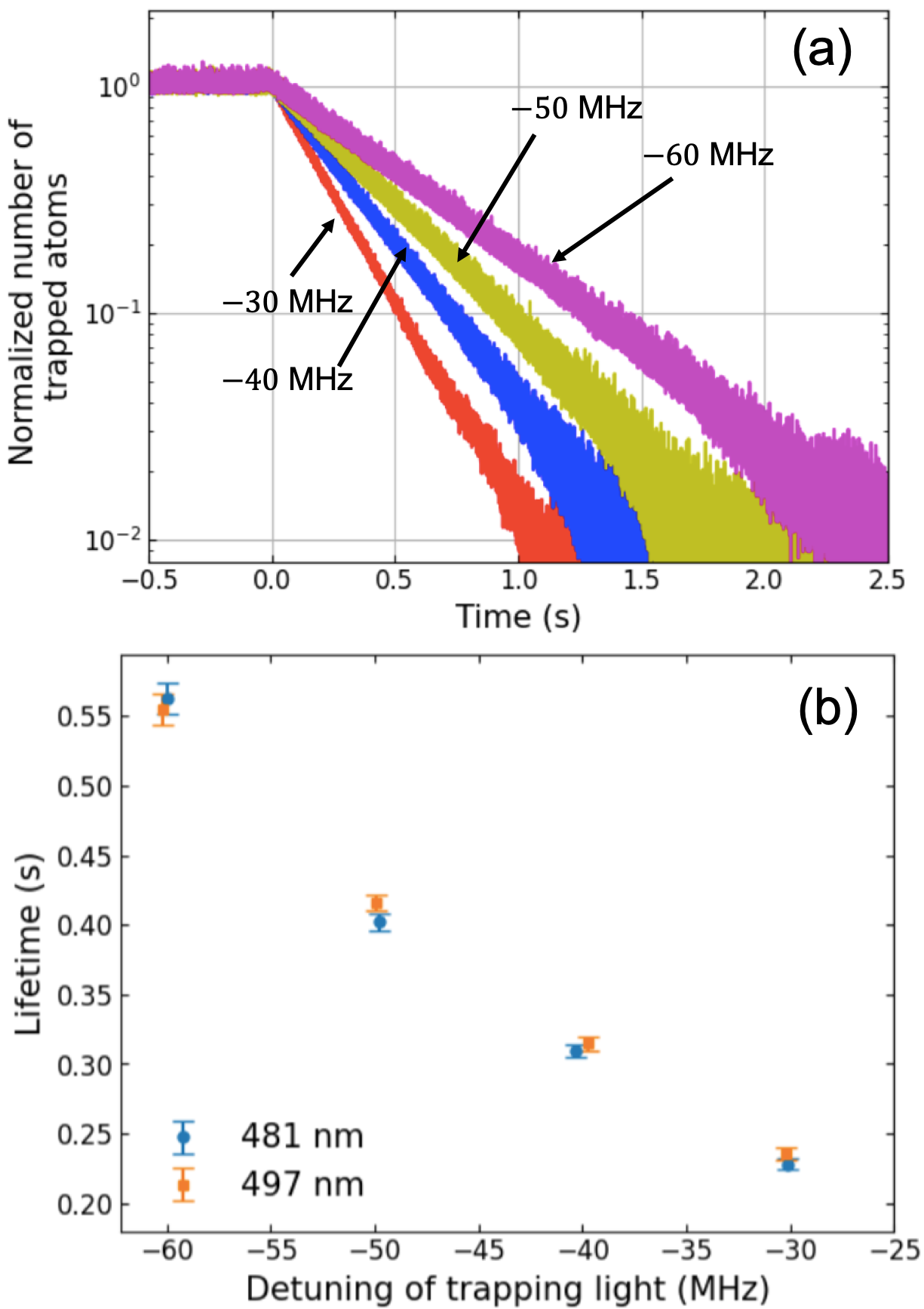}
		\caption{(a) Normalized number of trapped atoms after turning off the ${}^3P_0$ repumping laser ($483\,\mathrm{nm}$) while maintaining the ${}^3P_2$ repumping laser ($481\,\mathrm{nm}$) and varying the trapping light ($461\,\mathrm{nm}$) detuning. (b) Variation in lifetime with trapping light detuning for the two ${}^3P_2$ repumping lasers operating at $481\,\mathrm{nm}$ and $497\,\mathrm{nm}$.}
		\label{lifetime_combined}
	\end{center}
\end{figure}

\begin{figure}
	\begin{center}
		\includegraphics[width=86mm]{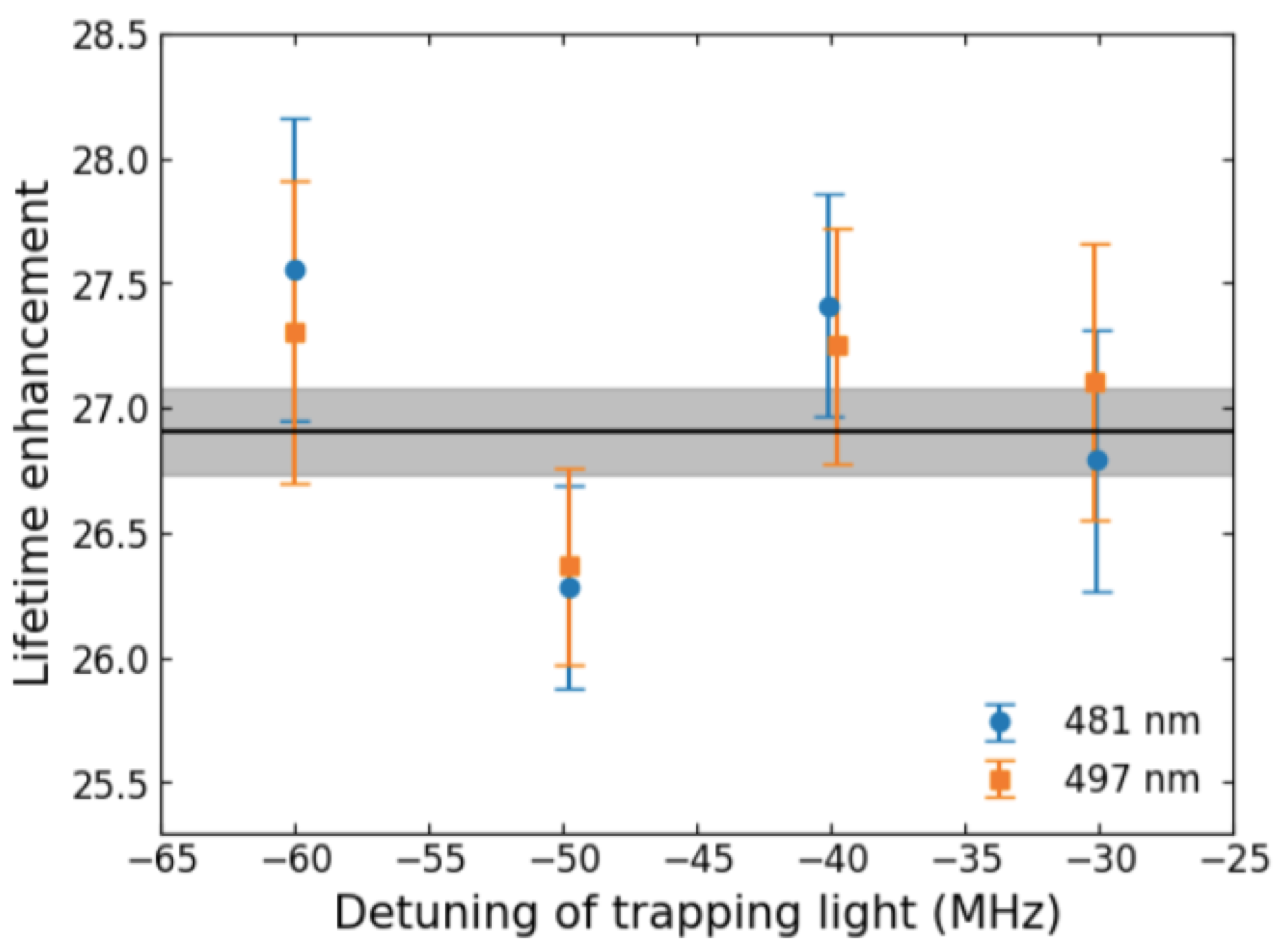}
		\caption{Enhancement in the lifetimes of the MOT for two single-repumping schemes operating at $481\,\mathrm{nm}$ and $497\,\mathrm{nm}$. The solid line is the weighted average of the data. The gray band represents the overall $\pm 1\sigma$ statistical uncertainty. Error bars indicate the uncertainty in fitting the decay data, as depicted in Fig.{\,}\ref{lifetime_combined}(a). The reduced $\chi^{2}$ value for the combined data is 1.1, with a $P$ value of 0.3.}
		\label{ratio_of_lifetime}
	\end{center}
\end{figure}

To estimate the decay rate for the transition from the $5s5p \,{}^1P_1$ state to the $5s5p \,{}^3P_0$ state for the two single-repumping schemes, we initially load the MOT using both the ${}^3P_0$ ($483\,\mathrm{nm}$) and ${}^3P_2$ ($481\,\mathrm{nm}$ or $497\,\mathrm{nm}$) repumping laser beams, followed by turning off the ${}^3P_0$ repumping laser beam. 
Figure \ref{lifetime_combined}(a) shows the decay of atom number in the MOT after turning off the 483-nm repumping light at various detunings of the trapping light ($461\,\mathrm{nm}$) when utilizing the 481-nm light. 
Figure \ref{lifetime_combined}(b) shows the dependence of the lifetime on the detuning of the trapping light for both the 481-nm and 497-nm repumping laser beams, indicating that the decay rate for the transition from the $5s5p \,{}^1P_1$ state to the $5s5p \,{}^3P_0$ state is independent of the choice of the ${}^3P_2$ repumping transition. 
The MOT lifetime increases with larger detuning because the fraction of atoms in the $5s5p \,{}^1P_1$ state diminishes with detuning as outlined in Eqs.{\,}\eqref{eq:occupation} and \eqref{eq:tau_MOT3} in Appendix \ref{rate_equation}.

To assess the improvement in lifetime for the two single-repumping schemes, we also measure the decay rate for the transition from the $5s5p \,{}^1P_1$ state to the $5s5p \,{}^3P_2$ state by turning off the ${}^3P_2$ ($481\,\mathrm{nm}$ or $497\,\mathrm{nm}$) repumping light while maintaining the ${}^3P_0$ ($483\,\mathrm{nm}$) repumping light (see Appendix \ref{rate_equation}).
Figure \ref{ratio_of_lifetime} shows the enhancement, defined as the ratio of the MOT lifetimes with and without the single-repumping light ($481\,\mathrm{nm}$ or $497\,\mathrm{nm}$) for various detunings. 
This enhancement remains unaffected by both the single-repumping transition and the detuning of the trapping light.
The weighted average enhancement factor is $26.9(2)$, consistent with the previously reported value of $32(5)$ for the 497-nm repumping scheme \cite{S.Stellmer2014}. 
We note that Fig.{\,}\ref{ratio_of_lifetime} indicates the enhancement factor is independent of the density of the trapped atoms because the density significantly depends on the detuning as described in Ref.{\,}\cite{X.Xu2003}.

\begin{figure}
	\begin{center}
		\includegraphics[width=86mm]{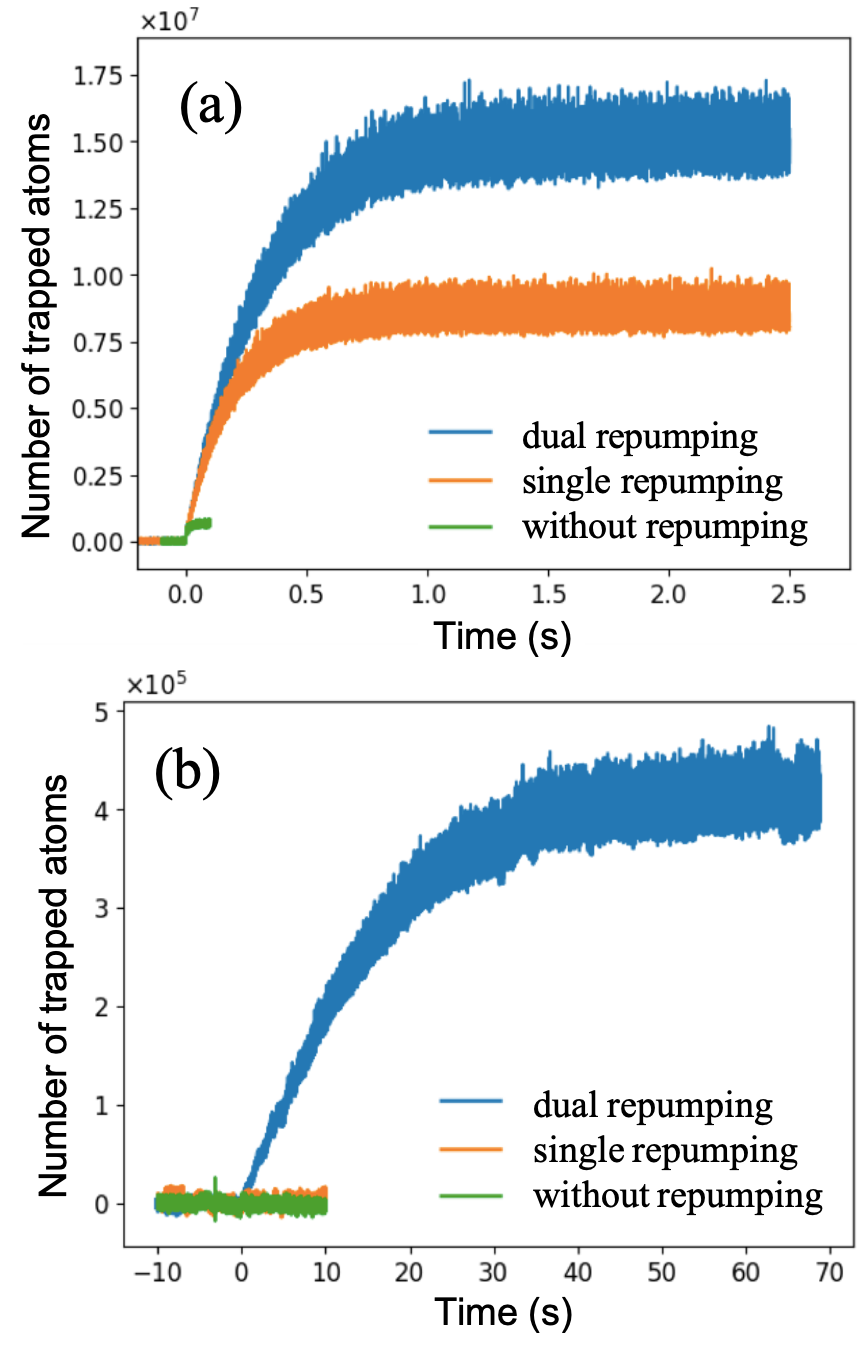}
		\caption{Evolution of the trapped atom number after turning on the MOT coils at oven temperatures of (a) $455\,\mathrm{{}^\circ C}$ and (b) $285\,\mathrm{{}^\circ C}$ under various repumping conditions: no repumping, single repumping operating at $481\,\mathrm{nm}$, and dual repumping operating at $481\,\mathrm{nm}$ and $483\,\mathrm{nm}$. The detuning of the trapping light is $-40\,\mathrm{MHz}$.}
		\label{oventemp_comparison}
	\end{center}
\end{figure}

The MOT lifetime, depicted in Fig.{\,}\ref{lifetime_combined}(a), using solely the ${}^3P_2$ ($481\,\mathrm{nm}$ or $497\,\mathrm{nm}$) repumping light, is less than $1\,\mathrm{s}$, considerably shorter compared with when both the ${}^3P_2$ and ${}^3P_0$ repumping laser beams are employed ($15\,\mathrm{s}$).
Previous studies employing a single-repumping scheme also consistently reported trap lifetimes of less than $1\,\mathrm{s}${\,}\cite{P.G.Mickelson2009, S.Stellmer2014, P.H.Moriya2018, F.Hu2019}. 
The decay from the upper state of the repumping transition to the ${}^3P_0$ state via intermediate states was mentioned{\,}\cite{S.Stellmer2014, C.Vishwakarma2019, P.H.Moriya2018, F.Hu2019, T.Akatsuka2021}.
If this hypothesis is the case, the enhancement factor should depend on the upper state (see Appendix \ref{rate_equation}).
However, as depicted in Fig.{\,}\ref{ratio_of_lifetime}, the enhancement factor remains unaffected by both the two single-repumping schemes, indicating that the decay from the upper state of the repumping transition to the ${}^3P_0$ state is negligible.
This is supported by the fact that, for the $497\,\mathrm{nm}$ repumping scheme, the branching ratio for the dipole-allowed transitions [$5s5d \,{}^3D_2 \to 5s6p \,{}^3P_2 \to 5s6s \,{}^3S_1\,(5s4d \,{}^3D_1) \to 5s5d \,{}^3P_0$] was estimated to be quite small ($0.035\%$){\,}\cite{S.Stellmer2014}, and, for the $481\,\mathrm{nm}$ repumping scheme, there is no intermediate states through which the upper state $5p^2  \,{}^3P_2$ is connected to the $5s5p \,{}^3P_0$ state by dipole-allowed transitions{\,}\cite{J.E.Sansonetti2010}.

As an alternative reason for the shorter MOT lifetimes for single-repumping schemes, binary collisional loss involving atoms in excited states was also mentioned{\,}\cite{F.Hu2019}. 
However, the single exponential decay observed in trapped atoms, as depicted in Fig.{\,}\ref{lifetime_combined}(a), rules out possible decay due to binary collisions. 
This is further supported by Fig.{\,}\ref{ratio_of_lifetime}, showing that the MOT lifetime enhancement is unaffected by detuning of the trapping light (i.e., MOT density). 

Our results strongly indicate, as discussed in Appendix \ref{estimation_decay}, that the primary decay path from the $5s5p \,{}^1P_1$ state to the $5s5p \,{}^3P_0$ state proceeds via the $5s4d \,{}^3D_1$ state as shown in Fig.{\,}\ref{Sr_energy_levels} ($5s5p \,{}^1P_1 \to 5s4d \,{}^3D_1 \to 5s5p \,{}^3P_0$). 
The branching ratio from the $5s5p \,{}^1P_1$ state (via the $5s4d \,{}^3D_1$ state) to the $5s5p \,{}^3P_0$ state is $1:3.9 \times 10^6$, derived from the branching ratio for the transition from the $5s5p \,{}^1P_1$ state (via the $5s4d \,{}^1D_2$ state) to the $5s5p \,{}^3P_2$ state ($1:150{\,}000$){\,}\cite{C.W.Bauschlicher1985, L.R.Hunter1986, X.Xu2003} and the observed enhancement factor of $26.9(2)$. 
Considering the branching ratio for the transition from the $5s4d \,{}^3D_1$ state to the $5s5p \,{}^3P_1$ state ($60\%$){\,}\cite{M.S.Safronova2013}, the decay rate for the transition from the $5s5p \,{}^1P_1$ state to the $5s4d \,{}^3D_1$ state is $83(32)\,\mathrm{s^{-1}}$ (see Appendix \ref{estimation_decay}); the uncertainty primarily arises from the decay rate for the transition from the $5s5p \,{}^1P_1$ state to the $5s4d \,{}^1D_2$ state [$3.9(1.5) \times 10^3\,\mathrm{s^{-1}}$]{\,}\cite{L.R.Hunter1986}.

Our results indicate that for a single-repumping scheme, if the loading rate of atoms is low enough to require a long loading time ($\gtrsim 1 \mathrm{s}$), the atom number in the MOT is considerably limited. 
This is experimentally demonstrated in Fig.{\,}\ref{oventemp_comparison}.
For instance, at an oven temperature of $455\,\mathrm{{}^\circ C}$ with a loading rate of $4 \times 10^{7}\,\mathrm{s^{-1}}$, the number of trapped atoms saturates within $1\,\mathrm{s}$ and the reduction in the number of trapped atoms for the single-repumping scheme is only $40\%$ compared with the dual-repumping scheme [Fig.{\,}\ref{oventemp_comparison}(a)]. 
Conversely, at an oven temperature of $285\,\mathrm{{}^\circ C}$ with a loading rate of $2 \times 10^{4}\,\mathrm{s^{-1}}$, the loading time of the MOT for the dual-repumping scheme exceeds $10\,\mathrm{s}$, and the number of trapped atoms is less than $10^4$ for the single-repumping scheme, being only 1/50 of that for the dual-repumping scheme [Fig.{\,}\ref{oventemp_comparison}(b)].

Considering the requirements for field-deployable or spaceborne optical lattice clocks{\,}\cite{S.B.Koller2017, J.Grotti2018, S.Origlia2018, W.Bowden2019, A.Sitaram2020, M.Takamoto2020, N.Ohmae2021, Y.B.Kale2022}, a small-sized and energy-saving atomic beam source is essential. 
Achieving these goals necessitates operating the atom oven at low temperatures and avoiding the use of Zeeman slowers and 2-dimensional MOTs{\,}\cite{I.Nosske2017, M.Barbiero2020}.
In this scenario, from which high MOT loading rates cannot be expected, a dual-repumping scheme should be adopted. 

\section{conclusion}
We conducted a comparative analysis of a MOT of $\mathrm{{}^{88} Sr}$ atoms for two single-repumping schemes ($481\,\mathrm{nm}$ and $497\,\mathrm{nm}$).
Our investigation revealed that the enhancement in lifetime was consistent at $26.9(2)$, demonstrating independence not only from the single-repumping transition but also from the density of trapped atoms. 
This outcome indicates that the primary decay path from the $5s5p \,{}^1P_1$ state to the $5s5p \,{}^3P_0$ state proceeds via the $5s4d \,{}^3D_1$ state ($5s5p \,{}^1P_1 \to 5s4d \,{}^3D_1 \to 5s5p \,{}^3P_0$), a mechanism previously overlooked. 
Owing to this decay path, the trapped atom number should be significantly restricted when employing a single-repumping scheme, particularly when a long loading time ($\gtrsim 1\,\mathrm{s}$) is required.
We estimated the branching ratio for the transition from the $5s5p \,{}^1P_1$ state (via the $5s4d \,{}^3D_1$ state) to the $5s5p \,{}^3P_0$ state to be $1:3.9 \times 10^6$ and the decay rate for the transition from the $5s5p \,{}^1P_1$ state to the $5s4d \,{}^3D_1$ state to be $83(32)\,\mathrm{s^{-1}}$. 
Our findings underscore the need to reevaluate single-repumping schemes for Sr laser cooling.

\section{acknowledgments}
We thank T. Sato and M. Mori for their contributions to the experiments.
This work was supported by JSPS KAKENHI Grant Numbers 23K20849 and 22KJ1163.

\appendix
\section{Rate equations}\label{rate_equation}
We present the rate equations describing the number of trapped atoms for various schemes:
\begin{enumerate}
    \item dual-repumping (${}^3P_2$ and ${}^3P_0$) scheme \label{case1}
    \item scheme with only the ${}^3P_0$ repumping laser \label{case2}
    \item single-repumping (${}^3P_2$) scheme \label{case3}
    \item scheme with no repumping laser \label{case4}
\end{enumerate}

For scheme \ref{case1}, the rate equation for the number of trapped atoms is given by
\begin{equation}
    \frac{dN}{dt}  = R - \gamma_c N - \beta N^2, \label{eq:MOT1}
\end{equation}
where $N$ denotes the number of trapped atoms, $R$ the atom loading rate, $\gamma_c$ the background collisional loss rate, and $\beta$ the coefficient for two-body collisional loss.
Under our experimental conditions (oven temperature of $335\,\mathrm{{}^\circ C}$), $\gamma_c$ and $\beta N$ are of the order of $0.1\,\mathrm{s^{-1}}$, confirmed by observing the decay of the number of trapped atoms after shutting off the atomic beam. 

For scheme \ref{case2}, in which atoms decaying to the ${}^3P_2$ state are lost, the rate equation is given by
\begin{equation}
    \frac{dN}{dt} = R - f a_2 N - \gamma_c N - \beta N^2, \label{eq:MOT2_temp1}
\end{equation}
where $a_{2}$ denotes the decay rate for the transition from the ${}^1P_1$ state to the ${}^3P_{2}$ state and $f$ the excitation fraction of the $5s^2 \,{}^1S_0 - 5s5p \,{}^1P_1$ ($461\,\mathrm{nm}$) transition, which is given by
\begin{equation}
    f =\frac{1}{2} \frac{s_0}{1+s_0+4(\Delta/\Gamma)^2}, \label{eq:occupation}
\end{equation}
where $s_0 = I/I_s$ ($I$ being the 461-nm laser beam intensity and $I_s = 40\,\mathrm{mW/cm^2}$ the saturation intensity) denotes the resonant saturation parameter, $\Delta$ the detuning, and $\Gamma = 2\pi\times 30\,\mathrm{MHz}$ the natural width.
According to Ref.{\,}\cite{X.Xu2003}, $a_2$ is given by
\begin{equation}
    a_2 = A_{{}^1P_1 \to {}^1D_2} B_{{}^1D_2 \to {}^3P_2}, \label{eq:a2} 
\end{equation}
where $A_{{}^1P_1 \to {}^1D_2} = 3.85(1.47)\times10^3\,\mathrm{s^{-1}}${\,}\cite{L.R.Hunter1986} denotes the decay rate for the transition from the ${}^1P_1$ state to the ${}^1D_2$ state, and $B_{{}^1D_2 \to {}^3P_2} = 0.33${\,}\cite{C.W.Bauschlicher1985} the branching ratio from the ${}^1D_2$ state to the ${}^3P_2$ state. 
Because the value of $f$ is typically $\sim 0.1$ for a standard MOT, $f a_2$ is $\sim 100\, \mathrm{s^{-1}}$, which is much larger than $\gamma_c$ and $\beta N$. 
Thus, Eq.{\,}\eqref{eq:MOT2_temp1} can be approximated by
\begin{equation}
    \frac{dN}{dt} = R - f a_2 N = R - \frac{N}{\tau_{2}}, \label{eq:MOT2}
\end{equation}
where $\tau_{2}$ is the lifetime of the MOT only with ${}^3P_0$ repumping laser, which is expressed as
\begin{equation}
    \tau_{2} = \frac{1}{f A_{{}^1P_1 \to {}^1D_2} B_{{}^1D_2 \to {}^3P_2}}. \label{eq:tau_483}
\end{equation}

For scheme \ref{case3}, in which atoms decaying to the ${}^3P_0$ state are lost, the rate equation is given by
\begin{equation}
    \frac{dN}{dt} = R - f a_{0}(X) N - \gamma_c N - \beta N^2, \label{eq:MOT3_temp}
\end{equation}
where $a_{0}(X)$ denotes the decay rate for the transition from the ${}^1P_1$ state to the ${}^3P_{0}$ state, and $X$ the upper state of the single-repumping transition ($X = 5p^2 \,{}^3P_2$ for $481\,\mathrm{nm}$ and $
X = 5s5d\,{}^3D_2$ for $497\,\mathrm{nm}$).
From Fig.{\,}\ref{lifetime_combined}(a), the observed decay rates of the MOT for the single-repumping scheme are of order $10\,\mathrm{s^{-1}}$, which is still much larger than $\gamma_c$ and $\beta N$.
Therefore, Eq.{\,}\eqref{eq:MOT3_temp} can be approximated by
\begin{equation}
    \frac{dN}{dt} = R - f a_{0}(X) N = R - \frac{N}{\tau_{0}(X)}, \label{eq:MOT3}
\end{equation}
where $\tau_{0}(X)$ is the MOT lifetime, which is expressed as
\begin{equation}
    \tau_{0}(X) = \frac{1}{f a_{0}(X)}. \label{eq:tau_MOT3}
\end{equation}

For scheme \ref{case4}, in which atoms decaying to the ${}^3P_0$ or ${}^3P_2$ state are lost, the rate equation is given by
\begin{equation}
    \frac{dN}{dt} = R - f (a_{0}(X) + a_{2}) N = R - \frac{N}{\tau_{\mathrm{norep}}(X)}, \label{eq:MOT4}
\end{equation}
for which terms $\gamma_c N$ and $\beta N^2$ are neglected, as in the scheme above, and $\tau_{\mathrm{norep}}(X)$ denotes the MOT lifetime, expressed as
\begin{equation}
    \tau_{\mathrm{norep}}(X) = \frac{1}{f (a_{0}(X)+a_2)}. \label{eq:tau_MOT4}
\end{equation}

To eliminate the uncertainty in the estimation of $f$, we measure the ratio $r (X) = \tau_{0}(X)/\tau_{2}$, which is expressed as
\begin{equation}
    r (X) = \frac{a_2}{a_{0}(X)}. \label{eq:ratio}
\end{equation}
The enhancement factor $\epsilon (X) = \tau_{0}(X)/\tau_{\mathrm{norep}}$ for the single-repumping scheme is then given by 
\begin{equation}
    \epsilon (X) = \frac{a_{0}(X)+a_2}{a_{0}(X)} = 1+r(X). \label{eq:epsilon}
\end{equation}

\section{Estimation of the decay rate from the $5s5p \,{}^1P_1$ state to the $5s4d \,{}^3D_1$ state}\label{estimation_decay}
The energy levels of the $5s4d \,{}^1D_2$ state and the $5s4d \,{}^3D_J$ state are below the $5s5p \,{}^1P_1$ state (see Fig.{\,}\ref{Sr_energy_levels}).
However, there is no electric-dipole allowed (E1) decay path from the $5s5p \,{}^1P_1$ state to the $5s5p \,{}^3P_0$ state through these states.
Our experimental results show the decay from the upper state of the $5s5p \,{}^3P_2$ repumping transition to the ${}^3P_0$ state via intermediate states is negligible. Therefore, we need to identify the decay path from the $5s5p \,{}^1P_1$ state to the $5s5p \,{}^3P_0$ state via the $5s4d \,{}^1D_2$ state or the $5s4d \,{}^3D_J$ state.
Considering that the decay rate for the spin-forbidden transition of $5s4d \,{}^1D_2 - 5s5p \,{}^3P_2$ ($1.9\,\mathrm{\mu m}$) is $661.1\,\mathrm{s^{-1}}$ (see Table.{\,}\ref{table:list} in Appendix \ref{tables}), it is reasonable to expect that the decay rates for the spin-forbidden transition of $5s5p \,{}^1P_1 - 5s4d \,{}^3D_1$ ($2.8\,\mathrm{\mu m}$) is of order $10^2 \sim 10^3\,\mathrm{s^{-1}}$.
Thus, we assume that the primary decay path from the $5s5p \,{}^1P_1$ state to the $5s5p \,{}^3P_0$ state is $5s5p \,{}^1P_1 \to 5s4d \,{}^3D_1 \to 5s5p \,{}^3P_0$.

Under this assumption, $a_0$ can be expressed as
\begin{equation}
    a_0 = A_{{}^1P_1 \to {}^3D_1} B_{{}^3D_1 \to {}^3P_0}, \label{eq:a0}
\end{equation}
where $A_{{}^1P_1 \to {}^3D_1}$ denotes the decay rate for the transition from the ${}^1P_1$ state to the ${}^3D_1$ state, and $B_{{}^3D_1 \to {}^3P_0} = 0.595${\,}\cite{M.S.Safronova2013} the branching ratio from the ${}^3D_1$ state to the ${}^3P_0$ state. 
From Eqs.{\,}\eqref{eq:ratio} and \eqref{eq:a0}, $A_{{}^1P_1 \to {}^3D_1}$ is expressed as
\begin{equation}
    A_{{}^1P_1 \to {}^3D_1} = \frac{a_2}{r B_{{}^3D_1 \to {}^3P_0}} = \frac{A_{{}^1P_1 \to {}^1D_2} B_{{}^1D_2 \to {}^3P_2}}{r B_{{}^3D_1 \to {}^3P_0}}. \label{eq:decay_rate}
\end{equation}
All the constants on the right-hand side of Eq.{\,}\eqref{eq:decay_rate} are known from the previous studies and our experiments: $A_{{}^1P_1 \to {}^1D_2} = 3.85(1.47)\times10^3\,\mathrm{s^{-1}}${\,}\cite{L.R.Hunter1986}, $B_{{}^1D_2 \to {}^3P_2} = 0.33${\,}\cite{C.W.Bauschlicher1985}, $B_{{}^3D_1 \to {}^3P_0} = 0.595${\,}\cite{M.S.Safronova2013}, $r= \epsilon - 1 = 25.9(2)$ [see Eq.{\,}\eqref{eq:epsilon} and Fig.{\,}\ref{ratio_of_lifetime}].
Then, one can calculate the decay rate for the $5s5p \,{}^1P_1 - 5s4d \,{}^3D_1$ transition as $A_{{}^1P_1 \to {}^3D_1} = 83(32)\,\mathrm{s^{-1}}$.

\section{Estimation of the excitation fraction and the saturation parameter}\label{estimation_f}
In general, estimating the total intensity of the MOT beam by measuring the beam power is difficult because of the uncertainties in gauging the shape of the beam and the loss of beam power in various optical components. 
To circumvent this problem, we adopt a method based on the excitation fraction.

From Eq.{\,}\eqref{eq:tau_483}, the excitation fraction of the 461-nm transition is expressed as
\begin{equation}
    f = \frac{1}{\tau_2 A_{{}^1P_1 \to {}^1D_2} B_{{}^1D_2 \to {}^3P_2}}, \label{eq:f} 
\end{equation}
where $\tau_{2}$ denotes the MOT lifetime obtained only with a ${}^3P_0$ repumping laser; the values of $A_{{}^1P_1 \to {}^1D_2}$ and $B_{{}^1D_2 \to {}^3P_2}$ are found in Ref.{\,}\cite{L.R.Hunter1986} and \cite{C.W.Bauschlicher1985}, respectively (see Appendix \ref{tables}).
Using Eq.{\,}\eqref{eq:f}, we can estimate $f$ by measuring $\tau_{2}$ with a relative uncertainty of $40\%$, stemming mainly from that of $A_{{}^1P_1 \to {}^1D_2}$.
From Eq.{\,}\eqref{eq:occupation}, we can infer the resonant saturation parameter $s_0$, and hence the total intensity $I = s_0 I_s$, where $I_s = 40\,\mathrm{mW/cm^2}$ is the saturation intensity.

\section{List of the decay rates} \label{tables}
We list the decay rates for the transitions relevant to our work in Table. \ref{table:list} .

\begin{table}[h]
    \caption{Decay rates for transitions relevant to the present work.}
    \label{table:list}
    \centering
    \begin{tabular}{ccc}
    \hline\hline 
      Transition & Decay rate ($\mathrm{s^{-1}}$) & Ref. \\ \hline
      $5s^2 \,{}^1S_0 - 5s5p \,{}^1P_1$ & $1.900 (1)\times10^8$ & \cite{M.Yasuda2006} \\
      $5s^2 \,{}^1S_0 - 5s5p \,{}^3P_1$ & $4.7 (1)\times10^4$ & \cite{R.Drozdowski1997} \\
      $5s5p \,{}^1P_1 - 5s4d \,{}^1D_2$ & $3.85 (1.47)\times10^3$ & \cite{L.R.Hunter1986} \\
      $5s4d \,{}^1D_2 - 5s5p \,{}^3P_2$ & 661.1 & \cite{C.W.Bauschlicher1985} \\
      $5s4d \,{}^1D_2 - 5s5p \,{}^3P_1$ & 1344.0 & \cite{C.W.Bauschlicher1985} \\
      $5s4d \,{}^3D_1 - 5s5p \,{}^3P_2$ & $8.510\times10^3$ & \cite{M.S.Safronova2013} \\
      $5s4d \,{}^3D_1 - 5s5p \,{}^3P_1$ & $1.777\times10^5$ & \cite{M.S.Safronova2013} \\
      $5s4d \,{}^3D_1 - 5s5p \,{}^3P_0$ & $2.740\times10^5$ & \cite{M.S.Safronova2013} \\
      $5s5p \,{}^3P_2 - 5p^2 \,{}^3P_2$ & $9.0 (6)\times10^7$ & \cite{J.E.Sansonetti2010} \\
      $5s5p \,{}^3P_2 - 5s5d \,{}^3D_2$ & $1.28 (9)\times10^7$ & \cite{J.E.Sansonetti2010} \\
      $5s5p \,{}^3P_0 - 5s5d \,{}^3D_1$ & $3.3 (2)\times10^7$ & \cite{J.E.Sansonetti2010} \\
      \hline\hline
    \end{tabular}
\end{table}

\bibliography{references.bib}

\begin{thebibliography}{53}%
\makeatletter
\providecommand \@ifxundefined [1]{%
 \@ifx{#1\undefined}
}%
\providecommand \@ifnum [1]{%
 \ifnum #1\expandafter \@firstoftwo
 \else \expandafter \@secondoftwo
 \fi
}%
\providecommand \@ifx [1]{%
 \ifx #1\expandafter \@firstoftwo
 \else \expandafter \@secondoftwo
 \fi
}%
\providecommand \natexlab [1]{#1}%
\providecommand \enquote  [1]{``#1''}%
\providecommand \bibnamefont  [1]{#1}%
\providecommand \bibfnamefont [1]{#1}%
\providecommand \citenamefont [1]{#1}%
\providecommand \href@noop [0]{\@secondoftwo}%
\providecommand \href [0]{\begingroup \@sanitize@url \@href}%
\providecommand \@href[1]{\@@startlink{#1}\@@href}%
\providecommand \@@href[1]{\endgroup#1\@@endlink}%
\providecommand \@sanitize@url [0]{\catcode `\\12\catcode `\$12\catcode `\&12\catcode `\#12\catcode `\^12\catcode `\_12\catcode `\%12\relax}%
\providecommand \@@startlink[1]{}%
\providecommand \@@endlink[0]{}%
\providecommand \url  [0]{\begingroup\@sanitize@url \@url }%
\providecommand \@url [1]{\endgroup\@href {#1}{\urlprefix }}%
\providecommand \urlprefix  [0]{URL }%
\providecommand \Eprint [0]{\href }%
\providecommand \doibase [0]{https://doi.org/}%
\providecommand \selectlanguage [0]{\@gobble}%
\providecommand \bibinfo  [0]{\@secondoftwo}%
\providecommand \bibfield  [0]{\@secondoftwo}%
\providecommand \translation [1]{[#1]}%
\providecommand \BibitemOpen [0]{}%
\providecommand \bibitemStop [0]{}%
\providecommand \bibitemNoStop [0]{.\EOS\space}%
\providecommand \EOS [0]{\spacefactor3000\relax}%
\providecommand \BibitemShut  [1]{\csname bibitem#1\endcsname}%
\let\auto@bib@innerbib\@empty
\bibitem [{\citenamefont {Campbell}\ \emph {et~al.}(2017)\citenamefont {Campbell}, \citenamefont {Hutson}, \citenamefont {Marti}, \citenamefont {Goban}, \citenamefont {Oppong}, \citenamefont {McNally}, \citenamefont {Sonderhouse}, \citenamefont {Robinson}, \citenamefont {Zhang}, \citenamefont {Bloom},\ and\ \citenamefont {Ye}}]{S.L.Campbell2017}%
  \BibitemOpen
  \bibfield  {author} {\bibinfo {author} {\bibfnamefont {S.~L.}\ \bibnamefont {Campbell}}, \bibinfo {author} {\bibfnamefont {R.~B.}\ \bibnamefont {Hutson}}, \bibinfo {author} {\bibfnamefont {G.~E.}\ \bibnamefont {Marti}}, \bibinfo {author} {\bibfnamefont {A.}~\bibnamefont {Goban}}, \bibinfo {author} {\bibfnamefont {N.~D.}\ \bibnamefont {Oppong}}, \bibinfo {author} {\bibfnamefont {R.~L.}\ \bibnamefont {McNally}}, \bibinfo {author} {\bibfnamefont {L.}~\bibnamefont {Sonderhouse}}, \bibinfo {author} {\bibfnamefont {J.~M.}\ \bibnamefont {Robinson}}, \bibinfo {author} {\bibfnamefont {W.}~\bibnamefont {Zhang}}, \bibinfo {author} {\bibfnamefont {B.~J.}\ \bibnamefont {Bloom}},\ and\ \bibinfo {author} {\bibfnamefont {J.}~\bibnamefont {Ye}},\ }\bibfield  {title} {\bibinfo {title} {A fermi-degenerate three-dimensional optical lattice clock},\ }\href {https://doi.org/10.1126/science.aam5538} {\bibfield  {journal} {\bibinfo  {journal} {Science}\ }\textbf {\bibinfo {volume} {358}},\ \bibinfo {pages} {90} (\bibinfo {year}
  {2017})}\BibitemShut {NoStop}%
\bibitem [{\citenamefont {Oelker}\ \emph {et~al.}(2019)\citenamefont {Oelker}, \citenamefont {Hutson}, \citenamefont {Kennedy}, \citenamefont {Sonderhouse}, \citenamefont {Bothwell}, \citenamefont {Goban}, \citenamefont {Kedar}, \citenamefont {Sanner}, \citenamefont {Robinson}, \citenamefont {Marti}, \citenamefont {Matei}, \citenamefont {Legero}, \citenamefont {Giunta}, \citenamefont {Holzwarth}, \citenamefont {Riehle}, \citenamefont {Sterr},\ and\ \citenamefont {Ye}}]{E.Oelker2019}%
  \BibitemOpen
  \bibfield  {author} {\bibinfo {author} {\bibfnamefont {E.}~\bibnamefont {Oelker}}, \bibinfo {author} {\bibfnamefont {R.~B.}\ \bibnamefont {Hutson}}, \bibinfo {author} {\bibfnamefont {C.~J.}\ \bibnamefont {Kennedy}}, \bibinfo {author} {\bibfnamefont {L.}~\bibnamefont {Sonderhouse}}, \bibinfo {author} {\bibfnamefont {T.}~\bibnamefont {Bothwell}}, \bibinfo {author} {\bibfnamefont {A.}~\bibnamefont {Goban}}, \bibinfo {author} {\bibfnamefont {D.}~\bibnamefont {Kedar}}, \bibinfo {author} {\bibfnamefont {C.}~\bibnamefont {Sanner}}, \bibinfo {author} {\bibfnamefont {J.~M.}\ \bibnamefont {Robinson}}, \bibinfo {author} {\bibfnamefont {G.~E.}\ \bibnamefont {Marti}}, \bibinfo {author} {\bibfnamefont {D.~G.}\ \bibnamefont {Matei}}, \bibinfo {author} {\bibfnamefont {T.}~\bibnamefont {Legero}}, \bibinfo {author} {\bibfnamefont {M.}~\bibnamefont {Giunta}}, \bibinfo {author} {\bibfnamefont {R.}~\bibnamefont {Holzwarth}}, \bibinfo {author} {\bibfnamefont {F.}~\bibnamefont {Riehle}}, \bibinfo {author} {\bibfnamefont
  {U.}~\bibnamefont {Sterr}},\ and\ \bibinfo {author} {\bibfnamefont {J.}~\bibnamefont {Ye}},\ }\bibfield  {title} {\bibinfo {title} {Demonstration of $4.8\times10^{-17}$ stability at 1{\thinspace}s for two independent optical clocks},\ }\href {https://doi.org/10.1038/s41566-019-0493-4} {\bibfield  {journal} {\bibinfo  {journal} {Nature Photonics}\ }\textbf {\bibinfo {volume} {13}},\ \bibinfo {pages} {714} (\bibinfo {year} {2019})}\BibitemShut {NoStop}%
\bibitem [{\citenamefont {Nicholson}\ \emph {et~al.}(2015)\citenamefont {Nicholson}, \citenamefont {Campbell}, \citenamefont {Hutson}, \citenamefont {Marti}, \citenamefont {Bloom}, \citenamefont {McNally}, \citenamefont {Zhang}, \citenamefont {Barrett}, \citenamefont {Safronova}, \citenamefont {Strouse}, \citenamefont {Tew},\ and\ \citenamefont {Ye}}]{T.L.Nicholson2015}%
  \BibitemOpen
  \bibfield  {author} {\bibinfo {author} {\bibfnamefont {T.~L.}\ \bibnamefont {Nicholson}}, \bibinfo {author} {\bibfnamefont {S.~L.}\ \bibnamefont {Campbell}}, \bibinfo {author} {\bibfnamefont {R.~B.}\ \bibnamefont {Hutson}}, \bibinfo {author} {\bibfnamefont {G.~E.}\ \bibnamefont {Marti}}, \bibinfo {author} {\bibfnamefont {B.~J.}\ \bibnamefont {Bloom}}, \bibinfo {author} {\bibfnamefont {R.~L.}\ \bibnamefont {McNally}}, \bibinfo {author} {\bibfnamefont {W.}~\bibnamefont {Zhang}}, \bibinfo {author} {\bibfnamefont {M.~D.}\ \bibnamefont {Barrett}}, \bibinfo {author} {\bibfnamefont {M.~S.}\ \bibnamefont {Safronova}}, \bibinfo {author} {\bibfnamefont {G.~F.}\ \bibnamefont {Strouse}}, \bibinfo {author} {\bibfnamefont {W.~L.}\ \bibnamefont {Tew}},\ and\ \bibinfo {author} {\bibfnamefont {J.}~\bibnamefont {Ye}},\ }\bibfield  {title} {\bibinfo {title} {Systematic evaluation of an atomic clock at $2 \times 10^{-18}$ total uncertainty},\ }\href {https://doi.org/10.1038/ncomms7896} {\bibfield  {journal} {\bibinfo
  {journal} {Nature Communications}\ }\textbf {\bibinfo {volume} {6}},\ \bibinfo {pages} {6896} (\bibinfo {year} {2015})}\BibitemShut {NoStop}%
\bibitem [{\citenamefont {McGrew}\ \emph {et~al.}(2018)\citenamefont {McGrew}, \citenamefont {Zhang}, \citenamefont {Fasano}, \citenamefont {Sch{\"a}ffer}, \citenamefont {Beloy}, \citenamefont {Nicolodi}, \citenamefont {Brown}, \citenamefont {Hinkley}, \citenamefont {Milani}, \citenamefont {Schioppo}, \citenamefont {Yoon},\ and\ \citenamefont {Ludlow}}]{W.F.McGrew2018}%
  \BibitemOpen
  \bibfield  {author} {\bibinfo {author} {\bibfnamefont {W.~F.}\ \bibnamefont {McGrew}}, \bibinfo {author} {\bibfnamefont {X.}~\bibnamefont {Zhang}}, \bibinfo {author} {\bibfnamefont {R.~J.}\ \bibnamefont {Fasano}}, \bibinfo {author} {\bibfnamefont {S.~A.}\ \bibnamefont {Sch{\"a}ffer}}, \bibinfo {author} {\bibfnamefont {K.}~\bibnamefont {Beloy}}, \bibinfo {author} {\bibfnamefont {D.}~\bibnamefont {Nicolodi}}, \bibinfo {author} {\bibfnamefont {R.~C.}\ \bibnamefont {Brown}}, \bibinfo {author} {\bibfnamefont {N.}~\bibnamefont {Hinkley}}, \bibinfo {author} {\bibfnamefont {G.}~\bibnamefont {Milani}}, \bibinfo {author} {\bibfnamefont {M.}~\bibnamefont {Schioppo}}, \bibinfo {author} {\bibfnamefont {T.~H.}\ \bibnamefont {Yoon}},\ and\ \bibinfo {author} {\bibfnamefont {A.~D.}\ \bibnamefont {Ludlow}},\ }\bibfield  {title} {\bibinfo {title} {Atomic clock performance enabling geodesy below the centimetre level},\ }\href {https://doi.org/10.1038/s41586-018-0738-2} {\bibfield  {journal} {\bibinfo  {journal} {Nature}\
  }\textbf {\bibinfo {volume} {564}},\ \bibinfo {pages} {87} (\bibinfo {year} {2018})}\BibitemShut {NoStop}%
\bibitem [{\citenamefont {Brewer}\ \emph {et~al.}(2019)\citenamefont {Brewer}, \citenamefont {Chen}, \citenamefont {Hankin}, \citenamefont {Clements}, \citenamefont {Chou}, \citenamefont {Wineland}, \citenamefont {Hume},\ and\ \citenamefont {Leibrandt}}]{S.M.Brewer2019}%
  \BibitemOpen
  \bibfield  {author} {\bibinfo {author} {\bibfnamefont {S.~M.}\ \bibnamefont {Brewer}}, \bibinfo {author} {\bibfnamefont {J.-S.}\ \bibnamefont {Chen}}, \bibinfo {author} {\bibfnamefont {A.~M.}\ \bibnamefont {Hankin}}, \bibinfo {author} {\bibfnamefont {E.~R.}\ \bibnamefont {Clements}}, \bibinfo {author} {\bibfnamefont {C.~W.}\ \bibnamefont {Chou}}, \bibinfo {author} {\bibfnamefont {D.~J.}\ \bibnamefont {Wineland}}, \bibinfo {author} {\bibfnamefont {D.~B.}\ \bibnamefont {Hume}},\ and\ \bibinfo {author} {\bibfnamefont {D.~R.}\ \bibnamefont {Leibrandt}},\ }\bibfield  {title} {\bibinfo {title} {$^{27}{\mathrm{al}}^{+}$ quantum-logic clock with a systematic uncertainty below ${10}^{\ensuremath{-}18}$},\ }\href {https://doi.org/10.1103/PhysRevLett.123.033201} {\bibfield  {journal} {\bibinfo  {journal} {Phys. Rev. Lett.}\ }\textbf {\bibinfo {volume} {123}},\ \bibinfo {pages} {033201} (\bibinfo {year} {2019})}\BibitemShut {NoStop}%
\bibitem [{\citenamefont {Bothwell}\ \emph {et~al.}(2019)\citenamefont {Bothwell}, \citenamefont {Kedar}, \citenamefont {Oelker}, \citenamefont {Robinson}, \citenamefont {Bromley}, \citenamefont {Tew}, \citenamefont {Ye},\ and\ \citenamefont {Kennedy}}]{T.Bothwell2019}%
  \BibitemOpen
  \bibfield  {author} {\bibinfo {author} {\bibfnamefont {T.}~\bibnamefont {Bothwell}}, \bibinfo {author} {\bibfnamefont {D.}~\bibnamefont {Kedar}}, \bibinfo {author} {\bibfnamefont {E.}~\bibnamefont {Oelker}}, \bibinfo {author} {\bibfnamefont {J.~M.}\ \bibnamefont {Robinson}}, \bibinfo {author} {\bibfnamefont {S.~L.}\ \bibnamefont {Bromley}}, \bibinfo {author} {\bibfnamefont {W.~L.}\ \bibnamefont {Tew}}, \bibinfo {author} {\bibfnamefont {J.}~\bibnamefont {Ye}},\ and\ \bibinfo {author} {\bibfnamefont {C.~J.}\ \bibnamefont {Kennedy}},\ }\bibfield  {title} {\bibinfo {title} {Jila sri optical lattice clock with uncertainty of $2.0 \times 10^{-18}$},\ }\href {https://doi.org/10.1088/1681-7575/ab4089} {\bibfield  {journal} {\bibinfo  {journal} {Metrologia}\ }\textbf {\bibinfo {volume} {56}},\ \bibinfo {pages} {065004} (\bibinfo {year} {2019})}\BibitemShut {NoStop}%
\bibitem [{\citenamefont {Nemitz}\ \emph {et~al.}(2016)\citenamefont {Nemitz}, \citenamefont {Ohkubo}, \citenamefont {Takamoto}, \citenamefont {Ushijima}, \citenamefont {Das}, \citenamefont {Ohmae},\ and\ \citenamefont {Katori}}]{N.Nemitz2016}%
  \BibitemOpen
  \bibfield  {author} {\bibinfo {author} {\bibfnamefont {N.}~\bibnamefont {Nemitz}}, \bibinfo {author} {\bibfnamefont {T.}~\bibnamefont {Ohkubo}}, \bibinfo {author} {\bibfnamefont {M.}~\bibnamefont {Takamoto}}, \bibinfo {author} {\bibfnamefont {I.}~\bibnamefont {Ushijima}}, \bibinfo {author} {\bibfnamefont {M.}~\bibnamefont {Das}}, \bibinfo {author} {\bibfnamefont {N.}~\bibnamefont {Ohmae}},\ and\ \bibinfo {author} {\bibfnamefont {H.}~\bibnamefont {Katori}},\ }\bibfield  {title} {\bibinfo {title} {Frequency ratio of yb and sr clocks with 5{\thinspace}{\texttimes}{\thinspace}$10^{-17}$ uncertainty at 150 seconds averaging time},\ }\href {https://doi.org/10.1038/nphoton.2016.20} {\bibfield  {journal} {\bibinfo  {journal} {Nature Photonics}\ }\textbf {\bibinfo {volume} {10}},\ \bibinfo {pages} {258} (\bibinfo {year} {2016})}\BibitemShut {NoStop}%
\bibitem [{\citenamefont {Beloy}\ \emph {et~al.}(2021)\citenamefont {Beloy}, \citenamefont {Bodine}, \citenamefont {Bothwell}, \citenamefont {Brewer}, \citenamefont {Bromley}, \citenamefont {Chen}, \citenamefont {Desch{\^e}nes}, \citenamefont {Diddams}, \citenamefont {Fasano}, \citenamefont {Fortier}, \citenamefont {Hassan}, \citenamefont {Hume}, \citenamefont {Kedar}, \citenamefont {Kennedy}, \citenamefont {Khader}, \citenamefont {Koepke}, \citenamefont {Leibrandt}, \citenamefont {Leopardi}, \citenamefont {Ludlow}, \citenamefont {McGrew}, \citenamefont {Milner}, \citenamefont {Newbury}, \citenamefont {Nicolodi}, \citenamefont {Oelker}, \citenamefont {Parker}, \citenamefont {Robinson}, \citenamefont {Romisch}, \citenamefont {Sch{\"a}ffer}, \citenamefont {Sherman}, \citenamefont {Sinclair}, \citenamefont {Sonderhouse}, \citenamefont {Swann}, \citenamefont {Yao}, \citenamefont {Ye}, \citenamefont {Zhang},\ and\ \citenamefont {Collaboration*}}]{BACONcolab2021}%
  \BibitemOpen
  \bibfield  {author} {\bibinfo {author} {\bibfnamefont {K.}~\bibnamefont {Beloy}}, \bibinfo {author} {\bibfnamefont {M.~I.}\ \bibnamefont {Bodine}}, \bibinfo {author} {\bibfnamefont {T.}~\bibnamefont {Bothwell}}, \bibinfo {author} {\bibfnamefont {S.~M.}\ \bibnamefont {Brewer}}, \bibinfo {author} {\bibfnamefont {S.~L.}\ \bibnamefont {Bromley}}, \bibinfo {author} {\bibfnamefont {J.-S.}\ \bibnamefont {Chen}}, \bibinfo {author} {\bibfnamefont {J.-D.}\ \bibnamefont {Desch{\^e}nes}}, \bibinfo {author} {\bibfnamefont {S.~A.}\ \bibnamefont {Diddams}}, \bibinfo {author} {\bibfnamefont {R.~J.}\ \bibnamefont {Fasano}}, \bibinfo {author} {\bibfnamefont {T.~M.}\ \bibnamefont {Fortier}}, \bibinfo {author} {\bibfnamefont {Y.~S.}\ \bibnamefont {Hassan}}, \bibinfo {author} {\bibfnamefont {D.~B.}\ \bibnamefont {Hume}}, \bibinfo {author} {\bibfnamefont {D.}~\bibnamefont {Kedar}}, \bibinfo {author} {\bibfnamefont {C.~J.}\ \bibnamefont {Kennedy}}, \bibinfo {author} {\bibfnamefont {I.}~\bibnamefont {Khader}}, \bibinfo {author}
  {\bibfnamefont {A.}~\bibnamefont {Koepke}}, \bibinfo {author} {\bibfnamefont {D.~R.}\ \bibnamefont {Leibrandt}}, \bibinfo {author} {\bibfnamefont {H.}~\bibnamefont {Leopardi}}, \bibinfo {author} {\bibfnamefont {A.~D.}\ \bibnamefont {Ludlow}}, \bibinfo {author} {\bibfnamefont {W.~F.}\ \bibnamefont {McGrew}}, \bibinfo {author} {\bibfnamefont {W.~R.}\ \bibnamefont {Milner}}, \bibinfo {author} {\bibfnamefont {N.~R.}\ \bibnamefont {Newbury}}, \bibinfo {author} {\bibfnamefont {D.}~\bibnamefont {Nicolodi}}, \bibinfo {author} {\bibfnamefont {E.}~\bibnamefont {Oelker}}, \bibinfo {author} {\bibfnamefont {T.~E.}\ \bibnamefont {Parker}}, \bibinfo {author} {\bibfnamefont {J.~M.}\ \bibnamefont {Robinson}}, \bibinfo {author} {\bibfnamefont {S.}~\bibnamefont {Romisch}}, \bibinfo {author} {\bibfnamefont {S.~A.}\ \bibnamefont {Sch{\"a}ffer}}, \bibinfo {author} {\bibfnamefont {J.~A.}\ \bibnamefont {Sherman}}, \bibinfo {author} {\bibfnamefont {L.~C.}\ \bibnamefont {Sinclair}}, \bibinfo {author} {\bibfnamefont {L.}~\bibnamefont
  {Sonderhouse}}, \bibinfo {author} {\bibfnamefont {W.~C.}\ \bibnamefont {Swann}}, \bibinfo {author} {\bibfnamefont {J.}~\bibnamefont {Yao}}, \bibinfo {author} {\bibfnamefont {J.}~\bibnamefont {Ye}}, \bibinfo {author} {\bibfnamefont {X.}~\bibnamefont {Zhang}},\ and\ \bibinfo {author} {\bibfnamefont {B.~A. C. O. N.~B.}\ \bibnamefont {Collaboration*}},\ }\bibfield  {title} {\bibinfo {title} {Frequency ratio measurements at 18-digit accuracy using an optical clock network},\ }\href {https://doi.org/10.1038/s41586-021-03253-4} {\bibfield  {journal} {\bibinfo  {journal} {Nature}\ }\textbf {\bibinfo {volume} {591}},\ \bibinfo {pages} {564} (\bibinfo {year} {2021})}\BibitemShut {NoStop}%
\bibitem [{\citenamefont {Delva}\ \emph {et~al.}(2017)\citenamefont {Delva}, \citenamefont {Lodewyck}, \citenamefont {Bilicki}, \citenamefont {Bookjans}, \citenamefont {Vallet}, \citenamefont {Le~Targat}, \citenamefont {Pottie}, \citenamefont {Guerlin}, \citenamefont {Meynadier}, \citenamefont {Le~Poncin-Lafitte}, \citenamefont {Lopez}, \citenamefont {Amy-Klein}, \citenamefont {Lee}, \citenamefont {Quintin}, \citenamefont {Lisdat}, \citenamefont {Al-Masoudi}, \citenamefont {D\"orscher}, \citenamefont {Grebing}, \citenamefont {Grosche}, \citenamefont {Kuhl}, \citenamefont {Raupach}, \citenamefont {Sterr}, \citenamefont {Hill}, \citenamefont {Hobson}, \citenamefont {Bowden}, \citenamefont {Kronj\"ager}, \citenamefont {Marra}, \citenamefont {Rolland}, \citenamefont {Baynes}, \citenamefont {Margolis},\ and\ \citenamefont {Gill}}]{P.Delva2017}%
  \BibitemOpen
  \bibfield  {author} {\bibinfo {author} {\bibfnamefont {P.}~\bibnamefont {Delva}}, \bibinfo {author} {\bibfnamefont {J.}~\bibnamefont {Lodewyck}}, \bibinfo {author} {\bibfnamefont {S.}~\bibnamefont {Bilicki}}, \bibinfo {author} {\bibfnamefont {E.}~\bibnamefont {Bookjans}}, \bibinfo {author} {\bibfnamefont {G.}~\bibnamefont {Vallet}}, \bibinfo {author} {\bibfnamefont {R.}~\bibnamefont {Le~Targat}}, \bibinfo {author} {\bibfnamefont {P.-E.}\ \bibnamefont {Pottie}}, \bibinfo {author} {\bibfnamefont {C.}~\bibnamefont {Guerlin}}, \bibinfo {author} {\bibfnamefont {F.}~\bibnamefont {Meynadier}}, \bibinfo {author} {\bibfnamefont {C.}~\bibnamefont {Le~Poncin-Lafitte}}, \bibinfo {author} {\bibfnamefont {O.}~\bibnamefont {Lopez}}, \bibinfo {author} {\bibfnamefont {A.}~\bibnamefont {Amy-Klein}}, \bibinfo {author} {\bibfnamefont {W.-K.}\ \bibnamefont {Lee}}, \bibinfo {author} {\bibfnamefont {N.}~\bibnamefont {Quintin}}, \bibinfo {author} {\bibfnamefont {C.}~\bibnamefont {Lisdat}}, \bibinfo {author} {\bibfnamefont
  {A.}~\bibnamefont {Al-Masoudi}}, \bibinfo {author} {\bibfnamefont {S.}~\bibnamefont {D\"orscher}}, \bibinfo {author} {\bibfnamefont {C.}~\bibnamefont {Grebing}}, \bibinfo {author} {\bibfnamefont {G.}~\bibnamefont {Grosche}}, \bibinfo {author} {\bibfnamefont {A.}~\bibnamefont {Kuhl}}, \bibinfo {author} {\bibfnamefont {S.}~\bibnamefont {Raupach}}, \bibinfo {author} {\bibfnamefont {U.}~\bibnamefont {Sterr}}, \bibinfo {author} {\bibfnamefont {I.~R.}\ \bibnamefont {Hill}}, \bibinfo {author} {\bibfnamefont {R.}~\bibnamefont {Hobson}}, \bibinfo {author} {\bibfnamefont {W.}~\bibnamefont {Bowden}}, \bibinfo {author} {\bibfnamefont {J.}~\bibnamefont {Kronj\"ager}}, \bibinfo {author} {\bibfnamefont {G.}~\bibnamefont {Marra}}, \bibinfo {author} {\bibfnamefont {A.}~\bibnamefont {Rolland}}, \bibinfo {author} {\bibfnamefont {F.~N.}\ \bibnamefont {Baynes}}, \bibinfo {author} {\bibfnamefont {H.~S.}\ \bibnamefont {Margolis}},\ and\ \bibinfo {author} {\bibfnamefont {P.}~\bibnamefont {Gill}},\ }\bibfield  {title} {\bibinfo
  {title} {Test of special relativity using a fiber network of optical clocks},\ }\href {https://doi.org/10.1103/PhysRevLett.118.221102} {\bibfield  {journal} {\bibinfo  {journal} {Phys. Rev. Lett.}\ }\textbf {\bibinfo {volume} {118}},\ \bibinfo {pages} {221102} (\bibinfo {year} {2017})}\BibitemShut {NoStop}%
\bibitem [{\citenamefont {Takamoto}\ \emph {et~al.}(2020)\citenamefont {Takamoto}, \citenamefont {Ushijima}, \citenamefont {Ohmae}, \citenamefont {Yahagi}, \citenamefont {Kokado}, \citenamefont {Shinkai},\ and\ \citenamefont {Katori}}]{M.Takamoto2020}%
  \BibitemOpen
  \bibfield  {author} {\bibinfo {author} {\bibfnamefont {M.}~\bibnamefont {Takamoto}}, \bibinfo {author} {\bibfnamefont {I.}~\bibnamefont {Ushijima}}, \bibinfo {author} {\bibfnamefont {N.}~\bibnamefont {Ohmae}}, \bibinfo {author} {\bibfnamefont {T.}~\bibnamefont {Yahagi}}, \bibinfo {author} {\bibfnamefont {K.}~\bibnamefont {Kokado}}, \bibinfo {author} {\bibfnamefont {H.}~\bibnamefont {Shinkai}},\ and\ \bibinfo {author} {\bibfnamefont {H.}~\bibnamefont {Katori}},\ }\bibfield  {title} {\bibinfo {title} {Test of general relativity by a pair of transportable optical lattice clocks},\ }\href {https://doi.org/10.1038/s41566-020-0619-8} {\bibfield  {journal} {\bibinfo  {journal} {Nature Photonics}\ }\textbf {\bibinfo {volume} {14}},\ \bibinfo {pages} {411} (\bibinfo {year} {2020})}\BibitemShut {NoStop}%
\bibitem [{\citenamefont {Bothwell}\ \emph {et~al.}(2022)\citenamefont {Bothwell}, \citenamefont {Kennedy}, \citenamefont {Aeppli}, \citenamefont {Kedar}, \citenamefont {Robinson}, \citenamefont {Oelker}, \citenamefont {Staron},\ and\ \citenamefont {Ye}}]{T.Bothwell2022}%
  \BibitemOpen
  \bibfield  {author} {\bibinfo {author} {\bibfnamefont {T.}~\bibnamefont {Bothwell}}, \bibinfo {author} {\bibfnamefont {C.~J.}\ \bibnamefont {Kennedy}}, \bibinfo {author} {\bibfnamefont {A.}~\bibnamefont {Aeppli}}, \bibinfo {author} {\bibfnamefont {D.}~\bibnamefont {Kedar}}, \bibinfo {author} {\bibfnamefont {J.~M.}\ \bibnamefont {Robinson}}, \bibinfo {author} {\bibfnamefont {E.}~\bibnamefont {Oelker}}, \bibinfo {author} {\bibfnamefont {A.}~\bibnamefont {Staron}},\ and\ \bibinfo {author} {\bibfnamefont {J.}~\bibnamefont {Ye}},\ }\bibfield  {title} {\bibinfo {title} {Resolving the gravitational redshift across a millimetre-scale atomic sample},\ }\href {https://doi.org/10.1038/s41586-021-04349-7} {\bibfield  {journal} {\bibinfo  {journal} {Nature}\ }\textbf {\bibinfo {volume} {602}},\ \bibinfo {pages} {420} (\bibinfo {year} {2022})}\BibitemShut {NoStop}%
\bibitem [{\citenamefont {Zheng}\ \emph {et~al.}(2023)\citenamefont {Zheng}, \citenamefont {Dolde}, \citenamefont {Cambria}, \citenamefont {Lim},\ and\ \citenamefont {Kolkowitz}}]{X.Zheng2023}%
  \BibitemOpen
  \bibfield  {author} {\bibinfo {author} {\bibfnamefont {X.}~\bibnamefont {Zheng}}, \bibinfo {author} {\bibfnamefont {J.}~\bibnamefont {Dolde}}, \bibinfo {author} {\bibfnamefont {M.~C.}\ \bibnamefont {Cambria}}, \bibinfo {author} {\bibfnamefont {H.~M.}\ \bibnamefont {Lim}},\ and\ \bibinfo {author} {\bibfnamefont {S.}~\bibnamefont {Kolkowitz}},\ }\bibfield  {title} {\bibinfo {title} {A lab-based test of the gravitational redshift with a miniature clock network},\ }\href {https://doi.org/10.1038/s41467-023-40629-8} {\bibfield  {journal} {\bibinfo  {journal} {Nature Communications}\ }\textbf {\bibinfo {volume} {14}},\ \bibinfo {pages} {4886} (\bibinfo {year} {2023})}\BibitemShut {NoStop}%
\bibitem [{\citenamefont {Kolkowitz}\ \emph {et~al.}(2017)\citenamefont {Kolkowitz}, \citenamefont {Bromley}, \citenamefont {Bothwell}, \citenamefont {Wall}, \citenamefont {Marti}, \citenamefont {Koller}, \citenamefont {Zhang}, \citenamefont {Rey},\ and\ \citenamefont {Ye}}]{S.Kolkowits2017}%
  \BibitemOpen
  \bibfield  {author} {\bibinfo {author} {\bibfnamefont {S.}~\bibnamefont {Kolkowitz}}, \bibinfo {author} {\bibfnamefont {S.~L.}\ \bibnamefont {Bromley}}, \bibinfo {author} {\bibfnamefont {T.}~\bibnamefont {Bothwell}}, \bibinfo {author} {\bibfnamefont {M.~L.}\ \bibnamefont {Wall}}, \bibinfo {author} {\bibfnamefont {G.~E.}\ \bibnamefont {Marti}}, \bibinfo {author} {\bibfnamefont {A.~P.}\ \bibnamefont {Koller}}, \bibinfo {author} {\bibfnamefont {X.}~\bibnamefont {Zhang}}, \bibinfo {author} {\bibfnamefont {A.~M.}\ \bibnamefont {Rey}},\ and\ \bibinfo {author} {\bibfnamefont {J.}~\bibnamefont {Ye}},\ }\bibfield  {title} {\bibinfo {title} {Spin--orbit-coupled fermions in an optical lattice clock},\ }\href {https://doi.org/10.1038/nature20811} {\bibfield  {journal} {\bibinfo  {journal} {Nature}\ }\textbf {\bibinfo {volume} {542}},\ \bibinfo {pages} {66} (\bibinfo {year} {2017})}\BibitemShut {NoStop}%
\bibitem [{\citenamefont {Norcia}\ \emph {et~al.}(2019)\citenamefont {Norcia}, \citenamefont {Young}, \citenamefont {Eckner}, \citenamefont {Oelker}, \citenamefont {Ye},\ and\ \citenamefont {Kaufman}}]{M.A.Norcia2019}%
  \BibitemOpen
  \bibfield  {author} {\bibinfo {author} {\bibfnamefont {M.~A.}\ \bibnamefont {Norcia}}, \bibinfo {author} {\bibfnamefont {A.~W.}\ \bibnamefont {Young}}, \bibinfo {author} {\bibfnamefont {W.~J.}\ \bibnamefont {Eckner}}, \bibinfo {author} {\bibfnamefont {E.}~\bibnamefont {Oelker}}, \bibinfo {author} {\bibfnamefont {J.}~\bibnamefont {Ye}},\ and\ \bibinfo {author} {\bibfnamefont {A.~M.}\ \bibnamefont {Kaufman}},\ }\bibfield  {title} {\bibinfo {title} {Seconds-scale coherence on an optical clock transition in a tweezer array},\ }\href {https://doi.org/10.1126/science.aay0644} {\bibfield  {journal} {\bibinfo  {journal} {Science}\ }\textbf {\bibinfo {volume} {366}},\ \bibinfo {pages} {93} (\bibinfo {year} {2019})}\BibitemShut {NoStop}%
\bibitem [{\citenamefont {Safronova}\ \emph {et~al.}(2018)\citenamefont {Safronova}, \citenamefont {Budker}, \citenamefont {DeMille}, \citenamefont {Kimball}, \citenamefont {Derevianko},\ and\ \citenamefont {Clark}}]{M.S.Safronova2018}%
  \BibitemOpen
  \bibfield  {author} {\bibinfo {author} {\bibfnamefont {M.~S.}\ \bibnamefont {Safronova}}, \bibinfo {author} {\bibfnamefont {D.}~\bibnamefont {Budker}}, \bibinfo {author} {\bibfnamefont {D.}~\bibnamefont {DeMille}}, \bibinfo {author} {\bibfnamefont {D.~F.~J.}\ \bibnamefont {Kimball}}, \bibinfo {author} {\bibfnamefont {A.}~\bibnamefont {Derevianko}},\ and\ \bibinfo {author} {\bibfnamefont {C.~W.}\ \bibnamefont {Clark}},\ }\bibfield  {title} {\bibinfo {title} {Search for new physics with atoms and molecules},\ }\href {https://doi.org/10.1103/RevModPhys.90.025008} {\bibfield  {journal} {\bibinfo  {journal} {Rev. Mod. Phys.}\ }\textbf {\bibinfo {volume} {90}},\ \bibinfo {pages} {025008} (\bibinfo {year} {2018})}\BibitemShut {NoStop}%
\bibitem [{\citenamefont {Kolkowitz}\ \emph {et~al.}(2016)\citenamefont {Kolkowitz}, \citenamefont {Pikovski}, \citenamefont {Langellier}, \citenamefont {Lukin}, \citenamefont {Walsworth},\ and\ \citenamefont {Ye}}]{S.Kolkowits2016}%
  \BibitemOpen
  \bibfield  {author} {\bibinfo {author} {\bibfnamefont {S.}~\bibnamefont {Kolkowitz}}, \bibinfo {author} {\bibfnamefont {I.}~\bibnamefont {Pikovski}}, \bibinfo {author} {\bibfnamefont {N.}~\bibnamefont {Langellier}}, \bibinfo {author} {\bibfnamefont {M.~D.}\ \bibnamefont {Lukin}}, \bibinfo {author} {\bibfnamefont {R.~L.}\ \bibnamefont {Walsworth}},\ and\ \bibinfo {author} {\bibfnamefont {J.}~\bibnamefont {Ye}},\ }\bibfield  {title} {\bibinfo {title} {Gravitational wave detection with optical lattice atomic clocks},\ }\href {https://doi.org/10.1103/PhysRevD.94.124043} {\bibfield  {journal} {\bibinfo  {journal} {Phys. Rev. D}\ }\textbf {\bibinfo {volume} {94}},\ \bibinfo {pages} {124043} (\bibinfo {year} {2016})}\BibitemShut {NoStop}%
\bibitem [{\citenamefont {Abe}\ \emph {et~al.}(2021)\citenamefont {Abe}, \citenamefont {Adamson}, \citenamefont {Borcean}, \citenamefont {Bortoletto}, \citenamefont {Bridges}, \citenamefont {Carman}, \citenamefont {Chattopadhyay}, \citenamefont {Coleman}, \citenamefont {Curfman}, \citenamefont {DeRose}, \citenamefont {Deshpande}, \citenamefont {Dimopoulos}, \citenamefont {Foot}, \citenamefont {Frisch}, \citenamefont {Garber}, \citenamefont {Geer}, \citenamefont {Gibson}, \citenamefont {Glick}, \citenamefont {Graham}, \citenamefont {Hahn}, \citenamefont {Harnik}, \citenamefont {Hawkins}, \citenamefont {Hindley}, \citenamefont {Hogan}, \citenamefont {Jiang}, \citenamefont {Kasevich}, \citenamefont {Kellett}, \citenamefont {Kiburg}, \citenamefont {Kovachy}, \citenamefont {Lykken}, \citenamefont {March-Russell}, \citenamefont {Mitchell}, \citenamefont {Murphy}, \citenamefont {Nantel}, \citenamefont {Nobrega}, \citenamefont {Plunkett}, \citenamefont {Rajendran}, \citenamefont {Rudolph}, \citenamefont {Sachdeva},
  \citenamefont {Safdari}, \citenamefont {Santucci}, \citenamefont {Schwartzman}, \citenamefont {Shipsey}, \citenamefont {Swan}, \citenamefont {Valerio}, \citenamefont {Vasonis}, \citenamefont {Wang},\ and\ \citenamefont {Wilkason}}]{M.Abe2021}%
  \BibitemOpen
  \bibfield  {author} {\bibinfo {author} {\bibfnamefont {M.}~\bibnamefont {Abe}}, \bibinfo {author} {\bibfnamefont {P.}~\bibnamefont {Adamson}}, \bibinfo {author} {\bibfnamefont {M.}~\bibnamefont {Borcean}}, \bibinfo {author} {\bibfnamefont {D.}~\bibnamefont {Bortoletto}}, \bibinfo {author} {\bibfnamefont {K.}~\bibnamefont {Bridges}}, \bibinfo {author} {\bibfnamefont {S.~P.}\ \bibnamefont {Carman}}, \bibinfo {author} {\bibfnamefont {S.}~\bibnamefont {Chattopadhyay}}, \bibinfo {author} {\bibfnamefont {J.}~\bibnamefont {Coleman}}, \bibinfo {author} {\bibfnamefont {N.~M.}\ \bibnamefont {Curfman}}, \bibinfo {author} {\bibfnamefont {K.}~\bibnamefont {DeRose}}, \bibinfo {author} {\bibfnamefont {T.}~\bibnamefont {Deshpande}}, \bibinfo {author} {\bibfnamefont {S.}~\bibnamefont {Dimopoulos}}, \bibinfo {author} {\bibfnamefont {C.~J.}\ \bibnamefont {Foot}}, \bibinfo {author} {\bibfnamefont {J.~C.}\ \bibnamefont {Frisch}}, \bibinfo {author} {\bibfnamefont {B.~E.}\ \bibnamefont {Garber}}, \bibinfo {author} {\bibfnamefont
  {S.}~\bibnamefont {Geer}}, \bibinfo {author} {\bibfnamefont {V.}~\bibnamefont {Gibson}}, \bibinfo {author} {\bibfnamefont {J.}~\bibnamefont {Glick}}, \bibinfo {author} {\bibfnamefont {P.~W.}\ \bibnamefont {Graham}}, \bibinfo {author} {\bibfnamefont {S.~R.}\ \bibnamefont {Hahn}}, \bibinfo {author} {\bibfnamefont {R.}~\bibnamefont {Harnik}}, \bibinfo {author} {\bibfnamefont {L.}~\bibnamefont {Hawkins}}, \bibinfo {author} {\bibfnamefont {S.}~\bibnamefont {Hindley}}, \bibinfo {author} {\bibfnamefont {J.~M.}\ \bibnamefont {Hogan}}, \bibinfo {author} {\bibfnamefont {Y.}~\bibnamefont {Jiang}}, \bibinfo {author} {\bibfnamefont {M.~A.}\ \bibnamefont {Kasevich}}, \bibinfo {author} {\bibfnamefont {R.~J.}\ \bibnamefont {Kellett}}, \bibinfo {author} {\bibfnamefont {M.}~\bibnamefont {Kiburg}}, \bibinfo {author} {\bibfnamefont {T.}~\bibnamefont {Kovachy}}, \bibinfo {author} {\bibfnamefont {J.~D.}\ \bibnamefont {Lykken}}, \bibinfo {author} {\bibfnamefont {J.}~\bibnamefont {March-Russell}}, \bibinfo {author} {\bibfnamefont
  {J.}~\bibnamefont {Mitchell}}, \bibinfo {author} {\bibfnamefont {M.}~\bibnamefont {Murphy}}, \bibinfo {author} {\bibfnamefont {M.}~\bibnamefont {Nantel}}, \bibinfo {author} {\bibfnamefont {L.~E.}\ \bibnamefont {Nobrega}}, \bibinfo {author} {\bibfnamefont {R.~K.}\ \bibnamefont {Plunkett}}, \bibinfo {author} {\bibfnamefont {S.}~\bibnamefont {Rajendran}}, \bibinfo {author} {\bibfnamefont {J.}~\bibnamefont {Rudolph}}, \bibinfo {author} {\bibfnamefont {N.}~\bibnamefont {Sachdeva}}, \bibinfo {author} {\bibfnamefont {M.}~\bibnamefont {Safdari}}, \bibinfo {author} {\bibfnamefont {J.~K.}\ \bibnamefont {Santucci}}, \bibinfo {author} {\bibfnamefont {A.~G.}\ \bibnamefont {Schwartzman}}, \bibinfo {author} {\bibfnamefont {I.}~\bibnamefont {Shipsey}}, \bibinfo {author} {\bibfnamefont {H.}~\bibnamefont {Swan}}, \bibinfo {author} {\bibfnamefont {L.~R.}\ \bibnamefont {Valerio}}, \bibinfo {author} {\bibfnamefont {A.}~\bibnamefont {Vasonis}}, \bibinfo {author} {\bibfnamefont {Y.}~\bibnamefont {Wang}},\ and\ \bibinfo {author}
  {\bibfnamefont {T.}~\bibnamefont {Wilkason}},\ }\bibfield  {title} {\bibinfo {title} {Matter-wave atomic gradiometer interferometric sensor (magis-100)},\ }\href {https://doi.org/10.1088/2058-9565/abf719} {\bibfield  {journal} {\bibinfo  {journal} {Quantum Science and Technology}\ }\textbf {\bibinfo {volume} {6}},\ \bibinfo {pages} {044003} (\bibinfo {year} {2021})}\BibitemShut {NoStop}%
\bibitem [{\citenamefont {Kobayashi}\ \emph {et~al.}(2022)\citenamefont {Kobayashi}, \citenamefont {Takamizawa}, \citenamefont {Akamatsu}, \citenamefont {Kawasaki}, \citenamefont {Nishiyama}, \citenamefont {Hosaka}, \citenamefont {Hisai}, \citenamefont {Wada}, \citenamefont {Inaba}, \citenamefont {Tanabe},\ and\ \citenamefont {Yasuda}}]{T.Kobayashi2022}%
  \BibitemOpen
  \bibfield  {author} {\bibinfo {author} {\bibfnamefont {T.}~\bibnamefont {Kobayashi}}, \bibinfo {author} {\bibfnamefont {A.}~\bibnamefont {Takamizawa}}, \bibinfo {author} {\bibfnamefont {D.}~\bibnamefont {Akamatsu}}, \bibinfo {author} {\bibfnamefont {A.}~\bibnamefont {Kawasaki}}, \bibinfo {author} {\bibfnamefont {A.}~\bibnamefont {Nishiyama}}, \bibinfo {author} {\bibfnamefont {K.}~\bibnamefont {Hosaka}}, \bibinfo {author} {\bibfnamefont {Y.}~\bibnamefont {Hisai}}, \bibinfo {author} {\bibfnamefont {M.}~\bibnamefont {Wada}}, \bibinfo {author} {\bibfnamefont {H.}~\bibnamefont {Inaba}}, \bibinfo {author} {\bibfnamefont {T.}~\bibnamefont {Tanabe}},\ and\ \bibinfo {author} {\bibfnamefont {M.}~\bibnamefont {Yasuda}},\ }\bibfield  {title} {\bibinfo {title} {Search for ultralight dark matter from long-term frequency comparisons of optical and microwave atomic clocks},\ }\href {https://doi.org/10.1103/PhysRevLett.129.241301} {\bibfield  {journal} {\bibinfo  {journal} {Phys. Rev. Lett.}\ }\textbf {\bibinfo {volume}
  {129}},\ \bibinfo {pages} {241301} (\bibinfo {year} {2022})}\BibitemShut {NoStop}%
\bibitem [{\citenamefont {Dimarcq}\ \emph {et~al.}(2024)\citenamefont {Dimarcq}, \citenamefont {Gertsvolf}, \citenamefont {Mileti}, \citenamefont {Bize}, \citenamefont {Oates}, \citenamefont {Peik}, \citenamefont {Calonico}, \citenamefont {Ido}, \citenamefont {Tavella}, \citenamefont {Meynadier}, \citenamefont {Petit}, \citenamefont {Panfilo}, \citenamefont {Bartholomew}, \citenamefont {Defraigne}, \citenamefont {Donley}, \citenamefont {Hedekvist}, \citenamefont {Sesia}, \citenamefont {Wouters}, \citenamefont {Dubé}, \citenamefont {Fang}, \citenamefont {Levi}, \citenamefont {Lodewyck}, \citenamefont {Margolis}, \citenamefont {Newell}, \citenamefont {Slyusarev}, \citenamefont {Weyers}, \citenamefont {Uzan}, \citenamefont {Yasuda}, \citenamefont {Yu}, \citenamefont {Rieck}, \citenamefont {Schnatz}, \citenamefont {Hanado}, \citenamefont {Fujieda}, \citenamefont {Pottie}, \citenamefont {Hanssen}, \citenamefont {Malimon},\ and\ \citenamefont {Ashby}}]{N.Dimarcq2024}%
  \BibitemOpen
  \bibfield  {author} {\bibinfo {author} {\bibfnamefont {N.}~\bibnamefont {Dimarcq}}, \bibinfo {author} {\bibfnamefont {M.}~\bibnamefont {Gertsvolf}}, \bibinfo {author} {\bibfnamefont {G.}~\bibnamefont {Mileti}}, \bibinfo {author} {\bibfnamefont {S.}~\bibnamefont {Bize}}, \bibinfo {author} {\bibfnamefont {C.~W.}\ \bibnamefont {Oates}}, \bibinfo {author} {\bibfnamefont {E.}~\bibnamefont {Peik}}, \bibinfo {author} {\bibfnamefont {D.}~\bibnamefont {Calonico}}, \bibinfo {author} {\bibfnamefont {T.}~\bibnamefont {Ido}}, \bibinfo {author} {\bibfnamefont {P.}~\bibnamefont {Tavella}}, \bibinfo {author} {\bibfnamefont {F.}~\bibnamefont {Meynadier}}, \bibinfo {author} {\bibfnamefont {G.}~\bibnamefont {Petit}}, \bibinfo {author} {\bibfnamefont {G.}~\bibnamefont {Panfilo}}, \bibinfo {author} {\bibfnamefont {J.}~\bibnamefont {Bartholomew}}, \bibinfo {author} {\bibfnamefont {P.}~\bibnamefont {Defraigne}}, \bibinfo {author} {\bibfnamefont {E.~A.}\ \bibnamefont {Donley}}, \bibinfo {author} {\bibfnamefont {P.~O.}\
  \bibnamefont {Hedekvist}}, \bibinfo {author} {\bibfnamefont {I.}~\bibnamefont {Sesia}}, \bibinfo {author} {\bibfnamefont {M.}~\bibnamefont {Wouters}}, \bibinfo {author} {\bibfnamefont {P.}~\bibnamefont {Dubé}}, \bibinfo {author} {\bibfnamefont {F.}~\bibnamefont {Fang}}, \bibinfo {author} {\bibfnamefont {F.}~\bibnamefont {Levi}}, \bibinfo {author} {\bibfnamefont {J.}~\bibnamefont {Lodewyck}}, \bibinfo {author} {\bibfnamefont {H.~S.}\ \bibnamefont {Margolis}}, \bibinfo {author} {\bibfnamefont {D.}~\bibnamefont {Newell}}, \bibinfo {author} {\bibfnamefont {S.}~\bibnamefont {Slyusarev}}, \bibinfo {author} {\bibfnamefont {S.}~\bibnamefont {Weyers}}, \bibinfo {author} {\bibfnamefont {J.-P.}\ \bibnamefont {Uzan}}, \bibinfo {author} {\bibfnamefont {M.}~\bibnamefont {Yasuda}}, \bibinfo {author} {\bibfnamefont {D.-H.}\ \bibnamefont {Yu}}, \bibinfo {author} {\bibfnamefont {C.}~\bibnamefont {Rieck}}, \bibinfo {author} {\bibfnamefont {H.}~\bibnamefont {Schnatz}}, \bibinfo {author} {\bibfnamefont {Y.}~\bibnamefont
  {Hanado}}, \bibinfo {author} {\bibfnamefont {M.}~\bibnamefont {Fujieda}}, \bibinfo {author} {\bibfnamefont {P.-E.}\ \bibnamefont {Pottie}}, \bibinfo {author} {\bibfnamefont {J.}~\bibnamefont {Hanssen}}, \bibinfo {author} {\bibfnamefont {A.}~\bibnamefont {Malimon}},\ and\ \bibinfo {author} {\bibfnamefont {N.}~\bibnamefont {Ashby}},\ }\bibfield  {title} {\bibinfo {title} {Roadmap towards the redefinition of the second},\ }\href {https://doi.org/10.1088/1681-7575/ad17d2} {\bibfield  {journal} {\bibinfo  {journal} {Metrologia}\ }\textbf {\bibinfo {volume} {61}},\ \bibinfo {pages} {012001} (\bibinfo {year} {2024})}\BibitemShut {NoStop}%
\bibitem [{\citenamefont {Hunter}\ \emph {et~al.}(1986)\citenamefont {Hunter}, \citenamefont {Walker},\ and\ \citenamefont {Weiss}}]{L.R.Hunter1986}%
  \BibitemOpen
  \bibfield  {author} {\bibinfo {author} {\bibfnamefont {L.~R.}\ \bibnamefont {Hunter}}, \bibinfo {author} {\bibfnamefont {W.~A.}\ \bibnamefont {Walker}},\ and\ \bibinfo {author} {\bibfnamefont {D.~S.}\ \bibnamefont {Weiss}},\ }\bibfield  {title} {\bibinfo {title} {Observation of an atomic stark--electric-quadrupole interference},\ }\href {https://doi.org/10.1103/PhysRevLett.56.823} {\bibfield  {journal} {\bibinfo  {journal} {Phys. Rev. Lett.}\ }\textbf {\bibinfo {volume} {56}},\ \bibinfo {pages} {823} (\bibinfo {year} {1986})}\BibitemShut {NoStop}%
\bibitem [{\citenamefont {Xu}\ \emph {et~al.}(2003)\citenamefont {Xu}, \citenamefont {Loftus}, \citenamefont {Hall}, \citenamefont {Gallagher},\ and\ \citenamefont {Ye}}]{X.Xu2003}%
  \BibitemOpen
  \bibfield  {author} {\bibinfo {author} {\bibfnamefont {X.}~\bibnamefont {Xu}}, \bibinfo {author} {\bibfnamefont {T.~H.}\ \bibnamefont {Loftus}}, \bibinfo {author} {\bibfnamefont {J.~L.}\ \bibnamefont {Hall}}, \bibinfo {author} {\bibfnamefont {A.}~\bibnamefont {Gallagher}},\ and\ \bibinfo {author} {\bibfnamefont {J.}~\bibnamefont {Ye}},\ }\bibfield  {title} {\bibinfo {title} {Cooling and trapping of atomic strontium},\ }\href {https://doi.org/10.1364/JOSAB.20.000968} {\bibfield  {journal} {\bibinfo  {journal} {J. Opt. Soc. Am. B}\ }\textbf {\bibinfo {volume} {20}},\ \bibinfo {pages} {968} (\bibinfo {year} {2003})}\BibitemShut {NoStop}%
\bibitem [{\citenamefont {Jr}\ \emph {et~al.}(1985)\citenamefont {Jr}, \citenamefont {Langhoff},\ and\ \citenamefont {Partridge}}]{C.W.Bauschlicher1985}%
  \BibitemOpen
  \bibfield  {author} {\bibinfo {author} {\bibfnamefont {C.~W.~B.}\ \bibnamefont {Jr}}, \bibinfo {author} {\bibfnamefont {S.~R.}\ \bibnamefont {Langhoff}},\ and\ \bibinfo {author} {\bibfnamefont {H.}~\bibnamefont {Partridge}},\ }\bibfield  {title} {\bibinfo {title} {The radiative lifetime of the 1d2 state of ca and sr: a core-valence treatment},\ }\href {https://doi.org/10.1088/0022-3700/18/8/011} {\bibfield  {journal} {\bibinfo  {journal} {Journal of Physics B: Atomic and Molecular Physics}\ }\textbf {\bibinfo {volume} {18}},\ \bibinfo {pages} {1523} (\bibinfo {year} {1985})}\BibitemShut {NoStop}%
\bibitem [{\citenamefont {Drozdowski}\ \emph {et~al.}(1997)\citenamefont {Drozdowski}, \citenamefont {Ignaciuk}, \citenamefont {Kwela},\ and\ \citenamefont {Heldt}}]{R.Drozdowski1997}%
  \BibitemOpen
  \bibfield  {author} {\bibinfo {author} {\bibfnamefont {R.}~\bibnamefont {Drozdowski}}, \bibinfo {author} {\bibfnamefont {M.}~\bibnamefont {Ignaciuk}}, \bibinfo {author} {\bibfnamefont {J.}~\bibnamefont {Kwela}},\ and\ \bibinfo {author} {\bibfnamefont {J.}~\bibnamefont {Heldt}},\ }\bibfield  {title} {\bibinfo {title} {Radiative lifetimes of the lowest 3p1 metastable states of ca and sr},\ }\href {https://doi.org/10.1007/s004600050300} {\bibfield  {journal} {\bibinfo  {journal} {Zeitschrift f{\"u}r Physik D Atoms, Molecules and Clusters}\ }\textbf {\bibinfo {volume} {41}},\ \bibinfo {pages} {125} (\bibinfo {year} {1997})}\BibitemShut {NoStop}%
\bibitem [{\citenamefont {Derevianko}(2001)}]{A.Derevianko2001}%
  \BibitemOpen
  \bibfield  {author} {\bibinfo {author} {\bibfnamefont {A.}~\bibnamefont {Derevianko}},\ }\bibfield  {title} {\bibinfo {title} {Feasibility of cooling and trapping metastable alkaline-earth atoms},\ }\href {https://doi.org/10.1103/PhysRevLett.87.023002} {\bibfield  {journal} {\bibinfo  {journal} {Phys. Rev. Lett.}\ }\textbf {\bibinfo {volume} {87}},\ \bibinfo {pages} {023002} (\bibinfo {year} {2001})}\BibitemShut {NoStop}%
\bibitem [{\citenamefont {Yasuda}\ and\ \citenamefont {Katori}(2004)}]{M.Yasuda2004}%
  \BibitemOpen
  \bibfield  {author} {\bibinfo {author} {\bibfnamefont {M.}~\bibnamefont {Yasuda}}\ and\ \bibinfo {author} {\bibfnamefont {H.}~\bibnamefont {Katori}},\ }\bibfield  {title} {\bibinfo {title} {Lifetime measurement of the $^{3}p_{2}$ metastable state of strontium atoms},\ }\href {https://doi.org/10.1103/PhysRevLett.92.153004} {\bibfield  {journal} {\bibinfo  {journal} {Phys. Rev. Lett.}\ }\textbf {\bibinfo {volume} {92}},\ \bibinfo {pages} {153004} (\bibinfo {year} {2004})}\BibitemShut {NoStop}%
\bibitem [{\citenamefont {Taichenachev}\ \emph {et~al.}(2006)\citenamefont {Taichenachev}, \citenamefont {Yudin}, \citenamefont {Oates}, \citenamefont {Hoyt}, \citenamefont {Barber},\ and\ \citenamefont {Hollberg}}]{A.V.Taichenachev2006}%
  \BibitemOpen
  \bibfield  {author} {\bibinfo {author} {\bibfnamefont {A.~V.}\ \bibnamefont {Taichenachev}}, \bibinfo {author} {\bibfnamefont {V.~I.}\ \bibnamefont {Yudin}}, \bibinfo {author} {\bibfnamefont {C.~W.}\ \bibnamefont {Oates}}, \bibinfo {author} {\bibfnamefont {C.~W.}\ \bibnamefont {Hoyt}}, \bibinfo {author} {\bibfnamefont {Z.~W.}\ \bibnamefont {Barber}},\ and\ \bibinfo {author} {\bibfnamefont {L.}~\bibnamefont {Hollberg}},\ }\bibfield  {title} {\bibinfo {title} {Magnetic field-induced spectroscopy of forbidden optical transitions with application to lattice-based optical atomic clocks},\ }\href {https://doi.org/10.1103/PhysRevLett.96.083001} {\bibfield  {journal} {\bibinfo  {journal} {Phys. Rev. Lett.}\ }\textbf {\bibinfo {volume} {96}},\ \bibinfo {pages} {083001} (\bibinfo {year} {2006})}\BibitemShut {NoStop}%
\bibitem [{\citenamefont {Barber}\ \emph {et~al.}(2006)\citenamefont {Barber}, \citenamefont {Hoyt}, \citenamefont {Oates}, \citenamefont {Hollberg}, \citenamefont {Taichenachev},\ and\ \citenamefont {Yudin}}]{Z.W.Barber2006}%
  \BibitemOpen
  \bibfield  {author} {\bibinfo {author} {\bibfnamefont {Z.~W.}\ \bibnamefont {Barber}}, \bibinfo {author} {\bibfnamefont {C.~W.}\ \bibnamefont {Hoyt}}, \bibinfo {author} {\bibfnamefont {C.~W.}\ \bibnamefont {Oates}}, \bibinfo {author} {\bibfnamefont {L.}~\bibnamefont {Hollberg}}, \bibinfo {author} {\bibfnamefont {A.~V.}\ \bibnamefont {Taichenachev}},\ and\ \bibinfo {author} {\bibfnamefont {V.~I.}\ \bibnamefont {Yudin}},\ }\bibfield  {title} {\bibinfo {title} {Direct excitation of the forbidden clock transition in neutral $^{174}\mathrm{Yb}$ atoms confined to an optical lattice},\ }\href {https://doi.org/10.1103/PhysRevLett.96.083002} {\bibfield  {journal} {\bibinfo  {journal} {Phys. Rev. Lett.}\ }\textbf {\bibinfo {volume} {96}},\ \bibinfo {pages} {083002} (\bibinfo {year} {2006})}\BibitemShut {NoStop}%
\bibitem [{\citenamefont {Vogel}(2002)}]{K.R.Vogel1999}%
  \BibitemOpen
  \bibfield  {author} {\bibinfo {author} {\bibfnamefont {K.}~\bibnamefont {Vogel}},\ }\emph {\bibinfo {title} {Laser cooling on a narrow atomic transition and measurement of the two body collision loss rate in a strontium magneto-optical trap}},\ \href@noop {} {Ph.D. thesis},\ \bibinfo  {school} {University of Corlado} (\bibinfo {year} {2002})\BibitemShut {NoStop}%
\bibitem [{\citenamefont {Poli}\ \emph {et~al.}(2005)\citenamefont {Poli}, \citenamefont {Drullinger}, \citenamefont {Ferrari}, \citenamefont {L\'eonard}, \citenamefont {Sorrentino},\ and\ \citenamefont {Tino}}]{N.Poli2005}%
  \BibitemOpen
  \bibfield  {author} {\bibinfo {author} {\bibfnamefont {N.}~\bibnamefont {Poli}}, \bibinfo {author} {\bibfnamefont {R.~E.}\ \bibnamefont {Drullinger}}, \bibinfo {author} {\bibfnamefont {G.}~\bibnamefont {Ferrari}}, \bibinfo {author} {\bibfnamefont {J.}~\bibnamefont {L\'eonard}}, \bibinfo {author} {\bibfnamefont {F.}~\bibnamefont {Sorrentino}},\ and\ \bibinfo {author} {\bibfnamefont {G.~M.}\ \bibnamefont {Tino}},\ }\bibfield  {title} {\bibinfo {title} {Cooling and trapping of ultracold strontium isotopic mixtures},\ }\href {https://doi.org/10.1103/PhysRevA.71.061403} {\bibfield  {journal} {\bibinfo  {journal} {Phys. Rev. A}\ }\textbf {\bibinfo {volume} {71}},\ \bibinfo {pages} {061403} (\bibinfo {year} {2005})}\BibitemShut {NoStop}%
\bibitem [{\citenamefont {Stellmer}\ and\ \citenamefont {Schreck}(2014)}]{S.Stellmer2014}%
  \BibitemOpen
  \bibfield  {author} {\bibinfo {author} {\bibfnamefont {S.}~\bibnamefont {Stellmer}}\ and\ \bibinfo {author} {\bibfnamefont {F.}~\bibnamefont {Schreck}},\ }\bibfield  {title} {\bibinfo {title} {Reservoir spectroscopy of $5s5p$ ${}^{3}p{}_{2}$--$5snd$ ${}^{3}{D}_{1,2,3}$ transitions in strontium},\ }\href {https://doi.org/10.1103/PhysRevA.90.022512} {\bibfield  {journal} {\bibinfo  {journal} {Phys. Rev. A}\ }\textbf {\bibinfo {volume} {90}},\ \bibinfo {pages} {022512} (\bibinfo {year} {2014})}\BibitemShut {NoStop}%
\bibitem [{\citenamefont {Hu}\ \emph {et~al.}(2019)\citenamefont {Hu}, \citenamefont {Nosske}, \citenamefont {Couturier}, \citenamefont {Tan}, \citenamefont {Qiao}, \citenamefont {Chen}, \citenamefont {Jiang}, \citenamefont {Zhu},\ and\ \citenamefont {Weidem\"uller}}]{F.Hu2019}%
  \BibitemOpen
  \bibfield  {author} {\bibinfo {author} {\bibfnamefont {F.}~\bibnamefont {Hu}}, \bibinfo {author} {\bibfnamefont {I.}~\bibnamefont {Nosske}}, \bibinfo {author} {\bibfnamefont {L.}~\bibnamefont {Couturier}}, \bibinfo {author} {\bibfnamefont {C.}~\bibnamefont {Tan}}, \bibinfo {author} {\bibfnamefont {C.}~\bibnamefont {Qiao}}, \bibinfo {author} {\bibfnamefont {P.}~\bibnamefont {Chen}}, \bibinfo {author} {\bibfnamefont {Y.~H.}\ \bibnamefont {Jiang}}, \bibinfo {author} {\bibfnamefont {B.}~\bibnamefont {Zhu}},\ and\ \bibinfo {author} {\bibfnamefont {M.}~\bibnamefont {Weidem\"uller}},\ }\bibfield  {title} {\bibinfo {title} {Analyzing a single-laser repumping scheme for efficient loading of a strontium magneto-optical trap},\ }\href {https://doi.org/10.1103/PhysRevA.99.033422} {\bibfield  {journal} {\bibinfo  {journal} {Phys. Rev. A}\ }\textbf {\bibinfo {volume} {99}},\ \bibinfo {pages} {033422} (\bibinfo {year} {2019})}\BibitemShut {NoStop}%
\bibitem [{\citenamefont {Mickelson}\ \emph {et~al.}(2009)\citenamefont {Mickelson}, \citenamefont {de~Escobar}, \citenamefont {Anzel}, \citenamefont {DeSalvo}, \citenamefont {Nagel}, \citenamefont {Traverso}, \citenamefont {Yan},\ and\ \citenamefont {Killian}}]{P.G.Mickelson2009}%
  \BibitemOpen
  \bibfield  {author} {\bibinfo {author} {\bibfnamefont {P.~G.}\ \bibnamefont {Mickelson}}, \bibinfo {author} {\bibfnamefont {Y.~N.~M.}\ \bibnamefont {de~Escobar}}, \bibinfo {author} {\bibfnamefont {P.}~\bibnamefont {Anzel}}, \bibinfo {author} {\bibfnamefont {B.~J.}\ \bibnamefont {DeSalvo}}, \bibinfo {author} {\bibfnamefont {S.~B.}\ \bibnamefont {Nagel}}, \bibinfo {author} {\bibfnamefont {A.~J.}\ \bibnamefont {Traverso}}, \bibinfo {author} {\bibfnamefont {M.}~\bibnamefont {Yan}},\ and\ \bibinfo {author} {\bibfnamefont {T.~C.}\ \bibnamefont {Killian}},\ }\bibfield  {title} {\bibinfo {title} {Repumping and spectroscopy of laser-cooled sr atoms using the (5s5p)3p2–(5s4d)3d2 transition},\ }\href {https://doi.org/10.1088/0953-4075/42/23/235001} {\bibfield  {journal} {\bibinfo  {journal} {Journal of Physics B: Atomic, Molecular and Optical Physics}\ }\textbf {\bibinfo {volume} {42}},\ \bibinfo {pages} {235001} (\bibinfo {year} {2009})}\BibitemShut {NoStop}%
\bibitem [{\citenamefont {Porsev}\ \emph {et~al.}(1999)\citenamefont {Porsev}, \citenamefont {Rakhlina},\ and\ \citenamefont {Kozlov}}]{S.G.Porsev1999}%
  \BibitemOpen
  \bibfield  {author} {\bibinfo {author} {\bibfnamefont {S.~G.}\ \bibnamefont {Porsev}}, \bibinfo {author} {\bibfnamefont {Y.~G.}\ \bibnamefont {Rakhlina}},\ and\ \bibinfo {author} {\bibfnamefont {M.~G.}\ \bibnamefont {Kozlov}},\ }\bibfield  {title} {\bibinfo {title} {Electric-dipole amplitudes, lifetimes, and polarizabilities of the low-lying levels of atomic ytterbium},\ }\href {https://doi.org/10.1103/PhysRevA.60.2781} {\bibfield  {journal} {\bibinfo  {journal} {Phys. Rev. A}\ }\textbf {\bibinfo {volume} {60}},\ \bibinfo {pages} {2781} (\bibinfo {year} {1999})}\BibitemShut {NoStop}%
\bibitem [{\citenamefont {Cho}\ \emph {et~al.}(2012)\citenamefont {Cho}, \citenamefont {Lee}, \citenamefont {Lee}, \citenamefont {Ahn}, \citenamefont {Lee}, \citenamefont {Yu}, \citenamefont {Lee},\ and\ \citenamefont {Park}}]{J.W.Cho2012}%
  \BibitemOpen
  \bibfield  {author} {\bibinfo {author} {\bibfnamefont {J.~W.}\ \bibnamefont {Cho}}, \bibinfo {author} {\bibfnamefont {H.-g.}\ \bibnamefont {Lee}}, \bibinfo {author} {\bibfnamefont {S.}~\bibnamefont {Lee}}, \bibinfo {author} {\bibfnamefont {J.}~\bibnamefont {Ahn}}, \bibinfo {author} {\bibfnamefont {W.-K.}\ \bibnamefont {Lee}}, \bibinfo {author} {\bibfnamefont {D.-H.}\ \bibnamefont {Yu}}, \bibinfo {author} {\bibfnamefont {S.~K.}\ \bibnamefont {Lee}},\ and\ \bibinfo {author} {\bibfnamefont {C.~Y.}\ \bibnamefont {Park}},\ }\bibfield  {title} {\bibinfo {title} {Optical repumping of triplet-$p$ states enhances magneto-optical trapping of ytterbium atoms},\ }\href {https://doi.org/10.1103/PhysRevA.85.035401} {\bibfield  {journal} {\bibinfo  {journal} {Phys. Rev. A}\ }\textbf {\bibinfo {volume} {85}},\ \bibinfo {pages} {035401} (\bibinfo {year} {2012})}\BibitemShut {NoStop}%
\bibitem [{\citenamefont {Sato}\ \emph {et~al.}(2022)\citenamefont {Sato}, \citenamefont {Hayakawa}, \citenamefont {Okamoto}, \citenamefont {Shimomura}, \citenamefont {Aoki},\ and\ \citenamefont {Torii}}]{T.Sato2022}%
  \BibitemOpen
  \bibfield  {author} {\bibinfo {author} {\bibfnamefont {T.}~\bibnamefont {Sato}}, \bibinfo {author} {\bibfnamefont {Y.}~\bibnamefont {Hayakawa}}, \bibinfo {author} {\bibfnamefont {N.}~\bibnamefont {Okamoto}}, \bibinfo {author} {\bibfnamefont {Y.}~\bibnamefont {Shimomura}}, \bibinfo {author} {\bibfnamefont {T.}~\bibnamefont {Aoki}},\ and\ \bibinfo {author} {\bibfnamefont {Y.}~\bibnamefont {Torii}},\ }\bibfield  {title} {\bibinfo {title} {Birefringent atomic vapor laser lock in a hollow cathode lamp},\ }\href {https://doi.org/10.1364/JOSAB.442465} {\bibfield  {journal} {\bibinfo  {journal} {J. Opt. Soc. Am. B}\ }\textbf {\bibinfo {volume} {39}},\ \bibinfo {pages} {155} (\bibinfo {year} {2022})}\BibitemShut {NoStop}%
\bibitem [{\citenamefont {Schioppo}\ \emph {et~al.}(2012)\citenamefont {Schioppo}, \citenamefont {Poli}, \citenamefont {Prevedelli}, \citenamefont {Falke}, \citenamefont {Lisdat}, \citenamefont {Sterr},\ and\ \citenamefont {Tino}}]{M.Schioppo2012}%
  \BibitemOpen
  \bibfield  {author} {\bibinfo {author} {\bibfnamefont {M.}~\bibnamefont {Schioppo}}, \bibinfo {author} {\bibfnamefont {N.}~\bibnamefont {Poli}}, \bibinfo {author} {\bibfnamefont {M.}~\bibnamefont {Prevedelli}}, \bibinfo {author} {\bibfnamefont {S.}~\bibnamefont {Falke}}, \bibinfo {author} {\bibfnamefont {C.}~\bibnamefont {Lisdat}}, \bibinfo {author} {\bibfnamefont {U.}~\bibnamefont {Sterr}},\ and\ \bibinfo {author} {\bibfnamefont {G.~M.}\ \bibnamefont {Tino}},\ }\bibfield  {title} {\bibinfo {title} {{A compact and efficient strontium oven for laser-cooling experiments}},\ }\href {https://doi.org/10.1063/1.4756936} {\bibfield  {journal} {\bibinfo  {journal} {Review of Scientific Instruments}\ }\textbf {\bibinfo {volume} {83}},\ \bibinfo {pages} {103101} (\bibinfo {year} {2012})}\BibitemShut {NoStop}%
\bibitem [{\citenamefont {Anderson}\ and\ \citenamefont {Kasevich}(1994)}]{B.P.Anderson1994}%
  \BibitemOpen
  \bibfield  {author} {\bibinfo {author} {\bibfnamefont {B.~P.}\ \bibnamefont {Anderson}}\ and\ \bibinfo {author} {\bibfnamefont {M.~A.}\ \bibnamefont {Kasevich}},\ }\bibfield  {title} {\bibinfo {title} {Enhanced loading of a magneto-optic trap from an atomic beam},\ }\href {https://doi.org/10.1103/PhysRevA.50.R3581} {\bibfield  {journal} {\bibinfo  {journal} {Phys. Rev. A}\ }\textbf {\bibinfo {volume} {50}},\ \bibinfo {pages} {R3581} (\bibinfo {year} {1994})}\BibitemShut {NoStop}%
\bibitem [{\citenamefont {Oates}\ \emph {et~al.}(1999)\citenamefont {Oates}, \citenamefont {Bondu}, \citenamefont {Fox},\ and\ \citenamefont {Hollberg}}]{C.W.Oates1999}%
  \BibitemOpen
  \bibfield  {author} {\bibinfo {author} {\bibfnamefont {C.~W.}\ \bibnamefont {Oates}}, \bibinfo {author} {\bibfnamefont {F.}~\bibnamefont {Bondu}}, \bibinfo {author} {\bibfnamefont {R.~W.}\ \bibnamefont {Fox}},\ and\ \bibinfo {author} {\bibfnamefont {L.}~\bibnamefont {Hollberg}},\ }\bibfield  {title} {\bibinfo {title} {A diode-laser optical frequency standard based on laser-cooled ca atoms: Sub-kilohertz spectroscopy by optical shelving detection},\ }\href {https://doi.org/10.1007/s100530050589} {\bibfield  {journal} {\bibinfo  {journal} {The European Physical Journal D - Atomic, Molecular, Optical and Plasma Physics}\ }\textbf {\bibinfo {volume} {7}},\ \bibinfo {pages} {449} (\bibinfo {year} {1999})}\BibitemShut {NoStop}%
\bibitem [{\citenamefont {Moriya}\ \emph {et~al.}(2018)\citenamefont {Moriya}, \citenamefont {Araújo}, \citenamefont {Todão}, \citenamefont {Hemmerling}, \citenamefont {Keßler}, \citenamefont {Shiozaki}, \citenamefont {Teixeira},\ and\ \citenamefont {Courteille}}]{P.H.Moriya2018}%
  \BibitemOpen
  \bibfield  {author} {\bibinfo {author} {\bibfnamefont {P.~H.}\ \bibnamefont {Moriya}}, \bibinfo {author} {\bibfnamefont {M.~O.}\ \bibnamefont {Araújo}}, \bibinfo {author} {\bibfnamefont {F.}~\bibnamefont {Todão}}, \bibinfo {author} {\bibfnamefont {M.}~\bibnamefont {Hemmerling}}, \bibinfo {author} {\bibfnamefont {H.}~\bibnamefont {Keßler}}, \bibinfo {author} {\bibfnamefont {R.~F.}\ \bibnamefont {Shiozaki}}, \bibinfo {author} {\bibfnamefont {R.~C.}\ \bibnamefont {Teixeira}},\ and\ \bibinfo {author} {\bibfnamefont {P.~W.}\ \bibnamefont {Courteille}},\ }\bibfield  {title} {\bibinfo {title} {Comparison between 403 nm and 497 nm repumping schemes for strontium magneto-optical traps},\ }\href {https://doi.org/10.1088/2399-6528/aaf662} {\bibfield  {journal} {\bibinfo  {journal} {Journal of Physics Communications}\ }\textbf {\bibinfo {volume} {2}},\ \bibinfo {pages} {125008} (\bibinfo {year} {2018})}\BibitemShut {NoStop}%
\bibitem [{\citenamefont {Vishwakarma}\ \emph {et~al.}(2019)\citenamefont {Vishwakarma}, \citenamefont {Patel}, \citenamefont {Mangaonkar}, \citenamefont {MacLennan}, \citenamefont {Biswas},\ and\ \citenamefont {Rapol}}]{C.Vishwakarma2019}%
  \BibitemOpen
  \bibfield  {author} {\bibinfo {author} {\bibfnamefont {C.}~\bibnamefont {Vishwakarma}}, \bibinfo {author} {\bibfnamefont {K.}~\bibnamefont {Patel}}, \bibinfo {author} {\bibfnamefont {J.}~\bibnamefont {Mangaonkar}}, \bibinfo {author} {\bibfnamefont {J.~L.}\ \bibnamefont {MacLennan}}, \bibinfo {author} {\bibfnamefont {K.}~\bibnamefont {Biswas}},\ and\ \bibinfo {author} {\bibfnamefont {U.~D.}\ \bibnamefont {Rapol}},\ }\href@noop {} {\bibinfo {title} {Study of loss dynamics of strontium in a magneto-optical trap}} (\bibinfo {year} {2019})\BibitemShut {NoStop}%
\bibitem [{\citenamefont {Akatsuka}\ \emph {et~al.}(2021)\citenamefont {Akatsuka}, \citenamefont {Hashiguchi}, \citenamefont {Takahashi}, \citenamefont {Ohmae}, \citenamefont {Takamoto},\ and\ \citenamefont {Katori}}]{T.Akatsuka2021}%
  \BibitemOpen
  \bibfield  {author} {\bibinfo {author} {\bibfnamefont {T.}~\bibnamefont {Akatsuka}}, \bibinfo {author} {\bibfnamefont {K.}~\bibnamefont {Hashiguchi}}, \bibinfo {author} {\bibfnamefont {T.}~\bibnamefont {Takahashi}}, \bibinfo {author} {\bibfnamefont {N.}~\bibnamefont {Ohmae}}, \bibinfo {author} {\bibfnamefont {M.}~\bibnamefont {Takamoto}},\ and\ \bibinfo {author} {\bibfnamefont {H.}~\bibnamefont {Katori}},\ }\bibfield  {title} {\bibinfo {title} {Three-stage laser cooling of sr atoms using the $5s5p^{3}p_{2}$ metastable state below doppler temperatures},\ }\href {https://doi.org/10.1103/PhysRevA.103.023331} {\bibfield  {journal} {\bibinfo  {journal} {Phys. Rev. A}\ }\textbf {\bibinfo {volume} {103}},\ \bibinfo {pages} {023331} (\bibinfo {year} {2021})}\BibitemShut {NoStop}%
\bibitem [{\citenamefont {Sansonetti}\ and\ \citenamefont {Nave}(2010)}]{J.E.Sansonetti2010}%
  \BibitemOpen
  \bibfield  {author} {\bibinfo {author} {\bibfnamefont {J.~E.}\ \bibnamefont {Sansonetti}}\ and\ \bibinfo {author} {\bibfnamefont {G.}~\bibnamefont {Nave}},\ }\bibfield  {title} {\bibinfo {title} {{Wavelengths, Transition Probabilities, and Energy Levels for the Spectrum of Neutral Strontium (SrI)}},\ }\href {https://doi.org/10.1063/1.3449176} {\bibfield  {journal} {\bibinfo  {journal} {Journal of Physical and Chemical Reference Data}\ }\textbf {\bibinfo {volume} {39}},\ \bibinfo {pages} {033103} (\bibinfo {year} {2010})}\BibitemShut {NoStop}%
\bibitem [{\citenamefont {Safronova}\ \emph {et~al.}(2013)\citenamefont {Safronova}, \citenamefont {Porsev}, \citenamefont {Safronova}, \citenamefont {Kozlov},\ and\ \citenamefont {Clark}}]{M.S.Safronova2013}%
  \BibitemOpen
  \bibfield  {author} {\bibinfo {author} {\bibfnamefont {M.~S.}\ \bibnamefont {Safronova}}, \bibinfo {author} {\bibfnamefont {S.~G.}\ \bibnamefont {Porsev}}, \bibinfo {author} {\bibfnamefont {U.~I.}\ \bibnamefont {Safronova}}, \bibinfo {author} {\bibfnamefont {M.~G.}\ \bibnamefont {Kozlov}},\ and\ \bibinfo {author} {\bibfnamefont {C.~W.}\ \bibnamefont {Clark}},\ }\bibfield  {title} {\bibinfo {title} {Blackbody-radiation shift in the sr optical atomic clock},\ }\href {https://doi.org/10.1103/PhysRevA.87.012509} {\bibfield  {journal} {\bibinfo  {journal} {Phys. Rev. A}\ }\textbf {\bibinfo {volume} {87}},\ \bibinfo {pages} {012509} (\bibinfo {year} {2013})}\BibitemShut {NoStop}%
\bibitem [{\citenamefont {Koller}\ \emph {et~al.}(2017)\citenamefont {Koller}, \citenamefont {Grotti}, \citenamefont {Vogt}, \citenamefont {Al-Masoudi}, \citenamefont {D\"orscher}, \citenamefont {H\"afner}, \citenamefont {Sterr},\ and\ \citenamefont {Lisdat}}]{S.B.Koller2017}%
  \BibitemOpen
  \bibfield  {author} {\bibinfo {author} {\bibfnamefont {S.~B.}\ \bibnamefont {Koller}}, \bibinfo {author} {\bibfnamefont {J.}~\bibnamefont {Grotti}}, \bibinfo {author} {\bibfnamefont {S.}~\bibnamefont {Vogt}}, \bibinfo {author} {\bibfnamefont {A.}~\bibnamefont {Al-Masoudi}}, \bibinfo {author} {\bibfnamefont {S.}~\bibnamefont {D\"orscher}}, \bibinfo {author} {\bibfnamefont {S.}~\bibnamefont {H\"afner}}, \bibinfo {author} {\bibfnamefont {U.}~\bibnamefont {Sterr}},\ and\ \bibinfo {author} {\bibfnamefont {C.}~\bibnamefont {Lisdat}},\ }\bibfield  {title} {\bibinfo {title} {Transportable optical lattice clock with $7\ifmmode\times\else\texttimes\fi{}{10}^{\ensuremath{-}17}$ uncertainty},\ }\href {https://doi.org/10.1103/PhysRevLett.118.073601} {\bibfield  {journal} {\bibinfo  {journal} {Phys. Rev. Lett.}\ }\textbf {\bibinfo {volume} {118}},\ \bibinfo {pages} {073601} (\bibinfo {year} {2017})}\BibitemShut {NoStop}%
\bibitem [{\citenamefont {Grotti}\ \emph {et~al.}(2018)\citenamefont {Grotti}, \citenamefont {Koller}, \citenamefont {Vogt}, \citenamefont {H{\"a}fner}, \citenamefont {Sterr}, \citenamefont {Lisdat}, \citenamefont {Denker}, \citenamefont {Voigt}, \citenamefont {Timmen}, \citenamefont {Rolland}, \citenamefont {Baynes}, \citenamefont {Margolis}, \citenamefont {Zampaolo}, \citenamefont {Thoumany}, \citenamefont {Pizzocaro}, \citenamefont {Rauf}, \citenamefont {Bregolin}, \citenamefont {Tampellini}, \citenamefont {Barbieri}, \citenamefont {Zucco}, \citenamefont {Costanzo}, \citenamefont {Clivati}, \citenamefont {Levi},\ and\ \citenamefont {Calonico}}]{J.Grotti2018}%
  \BibitemOpen
  \bibfield  {author} {\bibinfo {author} {\bibfnamefont {J.}~\bibnamefont {Grotti}}, \bibinfo {author} {\bibfnamefont {S.}~\bibnamefont {Koller}}, \bibinfo {author} {\bibfnamefont {S.}~\bibnamefont {Vogt}}, \bibinfo {author} {\bibfnamefont {S.}~\bibnamefont {H{\"a}fner}}, \bibinfo {author} {\bibfnamefont {U.}~\bibnamefont {Sterr}}, \bibinfo {author} {\bibfnamefont {C.}~\bibnamefont {Lisdat}}, \bibinfo {author} {\bibfnamefont {H.}~\bibnamefont {Denker}}, \bibinfo {author} {\bibfnamefont {C.}~\bibnamefont {Voigt}}, \bibinfo {author} {\bibfnamefont {L.}~\bibnamefont {Timmen}}, \bibinfo {author} {\bibfnamefont {A.}~\bibnamefont {Rolland}}, \bibinfo {author} {\bibfnamefont {F.~N.}\ \bibnamefont {Baynes}}, \bibinfo {author} {\bibfnamefont {H.~S.}\ \bibnamefont {Margolis}}, \bibinfo {author} {\bibfnamefont {M.}~\bibnamefont {Zampaolo}}, \bibinfo {author} {\bibfnamefont {P.}~\bibnamefont {Thoumany}}, \bibinfo {author} {\bibfnamefont {M.}~\bibnamefont {Pizzocaro}}, \bibinfo {author} {\bibfnamefont {B.}~\bibnamefont
  {Rauf}}, \bibinfo {author} {\bibfnamefont {F.}~\bibnamefont {Bregolin}}, \bibinfo {author} {\bibfnamefont {A.}~\bibnamefont {Tampellini}}, \bibinfo {author} {\bibfnamefont {P.}~\bibnamefont {Barbieri}}, \bibinfo {author} {\bibfnamefont {M.}~\bibnamefont {Zucco}}, \bibinfo {author} {\bibfnamefont {G.~A.}\ \bibnamefont {Costanzo}}, \bibinfo {author} {\bibfnamefont {C.}~\bibnamefont {Clivati}}, \bibinfo {author} {\bibfnamefont {F.}~\bibnamefont {Levi}},\ and\ \bibinfo {author} {\bibfnamefont {D.}~\bibnamefont {Calonico}},\ }\bibfield  {title} {\bibinfo {title} {Geodesy and metrology with a transportable optical clock},\ }\href {https://doi.org/10.1038/s41567-017-0042-3} {\bibfield  {journal} {\bibinfo  {journal} {Nature Physics}\ }\textbf {\bibinfo {volume} {14}},\ \bibinfo {pages} {437} (\bibinfo {year} {2018})}\BibitemShut {NoStop}%
\bibitem [{\citenamefont {Origlia}\ \emph {et~al.}(2018)\citenamefont {Origlia}, \citenamefont {Pramod}, \citenamefont {Schiller}, \citenamefont {Singh}, \citenamefont {Bongs}, \citenamefont {Schwarz}, \citenamefont {Al-Masoudi}, \citenamefont {D\"orscher}, \citenamefont {Herbers}, \citenamefont {H\"afner}, \citenamefont {Sterr},\ and\ \citenamefont {Lisdat}}]{S.Origlia2018}%
  \BibitemOpen
  \bibfield  {author} {\bibinfo {author} {\bibfnamefont {S.}~\bibnamefont {Origlia}}, \bibinfo {author} {\bibfnamefont {M.~S.}\ \bibnamefont {Pramod}}, \bibinfo {author} {\bibfnamefont {S.}~\bibnamefont {Schiller}}, \bibinfo {author} {\bibfnamefont {Y.}~\bibnamefont {Singh}}, \bibinfo {author} {\bibfnamefont {K.}~\bibnamefont {Bongs}}, \bibinfo {author} {\bibfnamefont {R.}~\bibnamefont {Schwarz}}, \bibinfo {author} {\bibfnamefont {A.}~\bibnamefont {Al-Masoudi}}, \bibinfo {author} {\bibfnamefont {S.}~\bibnamefont {D\"orscher}}, \bibinfo {author} {\bibfnamefont {S.}~\bibnamefont {Herbers}}, \bibinfo {author} {\bibfnamefont {S.}~\bibnamefont {H\"afner}}, \bibinfo {author} {\bibfnamefont {U.}~\bibnamefont {Sterr}},\ and\ \bibinfo {author} {\bibfnamefont {C.}~\bibnamefont {Lisdat}},\ }\bibfield  {title} {\bibinfo {title} {Towards an optical clock for space: Compact, high-performance optical lattice clock based on bosonic atoms},\ }\href {https://doi.org/10.1103/PhysRevA.98.053443} {\bibfield  {journal} {\bibinfo
  {journal} {Phys. Rev. A}\ }\textbf {\bibinfo {volume} {98}},\ \bibinfo {pages} {053443} (\bibinfo {year} {2018})}\BibitemShut {NoStop}%
\bibitem [{\citenamefont {Bowden}\ \emph {et~al.}(2019)\citenamefont {Bowden}, \citenamefont {Hobson}, \citenamefont {Hill}, \citenamefont {Vianello}, \citenamefont {Schioppo}, \citenamefont {Silva}, \citenamefont {Margolis}, \citenamefont {Baird},\ and\ \citenamefont {Gill}}]{W.Bowden2019}%
  \BibitemOpen
  \bibfield  {author} {\bibinfo {author} {\bibfnamefont {W.}~\bibnamefont {Bowden}}, \bibinfo {author} {\bibfnamefont {R.}~\bibnamefont {Hobson}}, \bibinfo {author} {\bibfnamefont {I.~R.}\ \bibnamefont {Hill}}, \bibinfo {author} {\bibfnamefont {A.}~\bibnamefont {Vianello}}, \bibinfo {author} {\bibfnamefont {M.}~\bibnamefont {Schioppo}}, \bibinfo {author} {\bibfnamefont {A.}~\bibnamefont {Silva}}, \bibinfo {author} {\bibfnamefont {H.~S.}\ \bibnamefont {Margolis}}, \bibinfo {author} {\bibfnamefont {P.~E.~G.}\ \bibnamefont {Baird}},\ and\ \bibinfo {author} {\bibfnamefont {P.}~\bibnamefont {Gill}},\ }\bibfield  {title} {\bibinfo {title} {A pyramid mot with integrated optical cavities as a cold atom platform for an optical lattice clock},\ }\href {https://doi.org/10.1038/s41598-019-48168-3} {\bibfield  {journal} {\bibinfo  {journal} {Scientific Reports}\ }\textbf {\bibinfo {volume} {9}},\ \bibinfo {pages} {11704} (\bibinfo {year} {2019})}\BibitemShut {NoStop}%
\bibitem [{\citenamefont {Sitaram}\ \emph {et~al.}(2020)\citenamefont {Sitaram}, \citenamefont {Elgee}, \citenamefont {Campbell}, \citenamefont {Klimov}, \citenamefont {Eckel},\ and\ \citenamefont {Barker}}]{A.Sitaram2020}%
  \BibitemOpen
  \bibfield  {author} {\bibinfo {author} {\bibfnamefont {A.}~\bibnamefont {Sitaram}}, \bibinfo {author} {\bibfnamefont {P.~K.}\ \bibnamefont {Elgee}}, \bibinfo {author} {\bibfnamefont {G.~K.}\ \bibnamefont {Campbell}}, \bibinfo {author} {\bibfnamefont {N.~N.}\ \bibnamefont {Klimov}}, \bibinfo {author} {\bibfnamefont {S.}~\bibnamefont {Eckel}},\ and\ \bibinfo {author} {\bibfnamefont {D.~S.}\ \bibnamefont {Barker}},\ }\bibfield  {title} {\bibinfo {title} {{Confinement of an alkaline-earth element in a grating magneto-optical trap}},\ }\href {https://doi.org/10.1063/5.0019551} {\bibfield  {journal} {\bibinfo  {journal} {Review of Scientific Instruments}\ }\textbf {\bibinfo {volume} {91}},\ \bibinfo {pages} {103202} (\bibinfo {year} {2020})}\BibitemShut {NoStop}%
\bibitem [{\citenamefont {Ohmae}\ \emph {et~al.}(2021)\citenamefont {Ohmae}, \citenamefont {Takamoto}, \citenamefont {Takahashi}, \citenamefont {Kokubun}, \citenamefont {Araki}, \citenamefont {Hinton}, \citenamefont {Ushijima}, \citenamefont {Muramatsu}, \citenamefont {Furumiya}, \citenamefont {Sakai}, \citenamefont {Moriya}, \citenamefont {Kamiya}, \citenamefont {Fujii}, \citenamefont {Muramatsu}, \citenamefont {Shiimado},\ and\ \citenamefont {Katori}}]{N.Ohmae2021}%
  \BibitemOpen
  \bibfield  {author} {\bibinfo {author} {\bibfnamefont {N.}~\bibnamefont {Ohmae}}, \bibinfo {author} {\bibfnamefont {M.}~\bibnamefont {Takamoto}}, \bibinfo {author} {\bibfnamefont {Y.}~\bibnamefont {Takahashi}}, \bibinfo {author} {\bibfnamefont {M.}~\bibnamefont {Kokubun}}, \bibinfo {author} {\bibfnamefont {K.}~\bibnamefont {Araki}}, \bibinfo {author} {\bibfnamefont {A.}~\bibnamefont {Hinton}}, \bibinfo {author} {\bibfnamefont {I.}~\bibnamefont {Ushijima}}, \bibinfo {author} {\bibfnamefont {T.}~\bibnamefont {Muramatsu}}, \bibinfo {author} {\bibfnamefont {T.}~\bibnamefont {Furumiya}}, \bibinfo {author} {\bibfnamefont {Y.}~\bibnamefont {Sakai}}, \bibinfo {author} {\bibfnamefont {N.}~\bibnamefont {Moriya}}, \bibinfo {author} {\bibfnamefont {N.}~\bibnamefont {Kamiya}}, \bibinfo {author} {\bibfnamefont {K.}~\bibnamefont {Fujii}}, \bibinfo {author} {\bibfnamefont {R.}~\bibnamefont {Muramatsu}}, \bibinfo {author} {\bibfnamefont {T.}~\bibnamefont {Shiimado}},\ and\ \bibinfo {author} {\bibfnamefont {H.}~\bibnamefont
  {Katori}},\ }\bibfield  {title} {\bibinfo {title} {Transportable strontium optical lattice clocks operated outside laboratory at the level of $10^{-18}$ uncertainty},\ }\href {https://doi.org/https://doi.org/10.1002/qute.202100015} {\bibfield  {journal} {\bibinfo  {journal} {Advanced Quantum Technologies}\ }\textbf {\bibinfo {volume} {4}},\ \bibinfo {pages} {2100015} (\bibinfo {year} {2021})}\BibitemShut {NoStop}%
\bibitem [{\citenamefont {Kale}\ \emph {et~al.}(2022)\citenamefont {Kale}, \citenamefont {Singh}, \citenamefont {Gellesch}, \citenamefont {Jones}, \citenamefont {Morris}, \citenamefont {Aldous}, \citenamefont {Bongs},\ and\ \citenamefont {Singh}}]{Y.B.Kale2022}%
  \BibitemOpen
  \bibfield  {author} {\bibinfo {author} {\bibfnamefont {Y.~B.}\ \bibnamefont {Kale}}, \bibinfo {author} {\bibfnamefont {A.}~\bibnamefont {Singh}}, \bibinfo {author} {\bibfnamefont {M.}~\bibnamefont {Gellesch}}, \bibinfo {author} {\bibfnamefont {J.~M.}\ \bibnamefont {Jones}}, \bibinfo {author} {\bibfnamefont {D.}~\bibnamefont {Morris}}, \bibinfo {author} {\bibfnamefont {M.}~\bibnamefont {Aldous}}, \bibinfo {author} {\bibfnamefont {K.}~\bibnamefont {Bongs}},\ and\ \bibinfo {author} {\bibfnamefont {Y.}~\bibnamefont {Singh}},\ }\bibfield  {title} {\bibinfo {title} {Field deployable atomics package for an optical lattice clock},\ }\href {https://doi.org/10.1088/2058-9565/ac7b40} {\bibfield  {journal} {\bibinfo  {journal} {Quantum Science and Technology}\ }\textbf {\bibinfo {volume} {7}},\ \bibinfo {pages} {045004} (\bibinfo {year} {2022})}\BibitemShut {NoStop}%
\bibitem [{\citenamefont {Nosske}\ \emph {et~al.}(2017)\citenamefont {Nosske}, \citenamefont {Couturier}, \citenamefont {Hu}, \citenamefont {Tan}, \citenamefont {Qiao}, \citenamefont {Blume}, \citenamefont {Jiang}, \citenamefont {Chen},\ and\ \citenamefont {Weidem\"uller}}]{I.Nosske2017}%
  \BibitemOpen
  \bibfield  {author} {\bibinfo {author} {\bibfnamefont {I.}~\bibnamefont {Nosske}}, \bibinfo {author} {\bibfnamefont {L.}~\bibnamefont {Couturier}}, \bibinfo {author} {\bibfnamefont {F.}~\bibnamefont {Hu}}, \bibinfo {author} {\bibfnamefont {C.}~\bibnamefont {Tan}}, \bibinfo {author} {\bibfnamefont {C.}~\bibnamefont {Qiao}}, \bibinfo {author} {\bibfnamefont {J.}~\bibnamefont {Blume}}, \bibinfo {author} {\bibfnamefont {Y.~H.}\ \bibnamefont {Jiang}}, \bibinfo {author} {\bibfnamefont {P.}~\bibnamefont {Chen}},\ and\ \bibinfo {author} {\bibfnamefont {M.}~\bibnamefont {Weidem\"uller}},\ }\bibfield  {title} {\bibinfo {title} {Two-dimensional magneto-optical trap as a source for cold strontium atoms},\ }\href {https://doi.org/10.1103/PhysRevA.96.053415} {\bibfield  {journal} {\bibinfo  {journal} {Phys. Rev. A}\ }\textbf {\bibinfo {volume} {96}},\ \bibinfo {pages} {053415} (\bibinfo {year} {2017})}\BibitemShut {NoStop}%
\bibitem [{\citenamefont {Barbiero}\ \emph {et~al.}(2020)\citenamefont {Barbiero}, \citenamefont {Tarallo}, \citenamefont {Calonico}, \citenamefont {Levi}, \citenamefont {Lamporesi},\ and\ \citenamefont {Ferrari}}]{M.Barbiero2020}%
  \BibitemOpen
  \bibfield  {author} {\bibinfo {author} {\bibfnamefont {M.}~\bibnamefont {Barbiero}}, \bibinfo {author} {\bibfnamefont {M.~G.}\ \bibnamefont {Tarallo}}, \bibinfo {author} {\bibfnamefont {D.}~\bibnamefont {Calonico}}, \bibinfo {author} {\bibfnamefont {F.}~\bibnamefont {Levi}}, \bibinfo {author} {\bibfnamefont {G.}~\bibnamefont {Lamporesi}},\ and\ \bibinfo {author} {\bibfnamefont {G.}~\bibnamefont {Ferrari}},\ }\bibfield  {title} {\bibinfo {title} {Sideband-enhanced cold atomic source for optical clocks},\ }\href {https://doi.org/10.1103/PhysRevApplied.13.014013} {\bibfield  {journal} {\bibinfo  {journal} {Phys. Rev. Appl.}\ }\textbf {\bibinfo {volume} {13}},\ \bibinfo {pages} {014013} (\bibinfo {year} {2020})}\BibitemShut {NoStop}%
\bibitem [{\citenamefont {Yasuda}\ \emph {et~al.}(2006)\citenamefont {Yasuda}, \citenamefont {Kishimoto}, \citenamefont {Takamoto},\ and\ \citenamefont {Katori}}]{M.Yasuda2006}%
  \BibitemOpen
  \bibfield  {author} {\bibinfo {author} {\bibfnamefont {M.}~\bibnamefont {Yasuda}}, \bibinfo {author} {\bibfnamefont {T.}~\bibnamefont {Kishimoto}}, \bibinfo {author} {\bibfnamefont {M.}~\bibnamefont {Takamoto}},\ and\ \bibinfo {author} {\bibfnamefont {H.}~\bibnamefont {Katori}},\ }\bibfield  {title} {\bibinfo {title} {Photoassociation spectroscopy of $^{88}\mathrm{Sr}$: Reconstruction of the wave function near the last node},\ }\href {https://doi.org/10.1103/PhysRevA.73.011403} {\bibfield  {journal} {\bibinfo  {journal} {Phys. Rev. A}\ }\textbf {\bibinfo {volume} {73}},\ \bibinfo {pages} {011403} (\bibinfo {year} {2006})}\BibitemShut {NoStop}%
\end{thebibliography}%

\end{document}